\documentclass[referee]{aa} 
\usepackage{graphicx}
\usepackage{epsfig}
\usepackage{array}
\epsfclipon
\def\kms{km~s$^{-1}$}
\def\he{HE 2243$-$6031}
\def\cm2{cm$^{-2}$}

\def\icm{cm$^{-2}$}
\def\lya{Ly$\alpha$}
\def\lyb{Ly$\beta$}

\def\fetwo{\ion{Fe}{ii}}
\def\fethree{\ion{Fe}{iii}}

\def\sitwo{\ion{Si}{ii}}

\def\sifour{\ion{Si}{iv}}

\def\oone{\ion{O}{vi}{ $\lambda1032$}}  
  
\def\cfour{\ion{C}{iv}}  
  
\def\ctwo{\ion{C}{ii}}  
\def\si4{\ion{Si}{iv}}  
\def\s6{\ion{S}{vi}}  
\def\o6{\ion{O}{vi}}  
\def\oone{\ion{O}{i}}  
\def\n5{\ion{N}{v}}  
\def\hone{\ion{H}{i}}  
  
\def\hetwo{\ion{He}{ii}}  
\def\altwo{\ion{Al}{ii}}  
\def\althree{\ion{Al}{iii}}  
\def\none{\ion{N}{i}}  
\def\ntwo{\ion{N}{ii}}  

\def\c43{\ion{C}{iv}/\ion{C}{iii}}

\def\ptwo{\ion{P}{ii}}  
\def\pthree{\ion{P}{iii}}  
\def\fetwo{\ion{Fe}{ii}}  
\def\nitwo{\ion{Ni}{ii}}  
\def\zntwo{\ion{Zn}{ii}}  
\def\mntwo{\ion{Mn}{ii}}  
\def\crtwo{\ion{Cr}{ii}}  
\def\stwo{\ion{S}{ii}}  
\def\sthree{\ion{S}{iii}}  
\def\arone{\ion{Ar}{i}}  
\def\artwo{\ion{Ar}{ii}}  

\begin{document}

\title{
Metal abundances and ionization conditions in a possibly dust-free damped
\lya\ system at $z=2.3$\thanks{Based on observations
collected at the    
European Southern Observatory, 
Chile (Programme ID: 65.O-0411(A))
}}

   \author{Sebastian Lopez
          \inst{1}
          \and
          Dieter Reimers\inst{2}
          \and
          Sandro D'Odorico\inst{3}
          \and
          {\bf Jason X. Prochaska}\inst{4,}\thanks{Hubble Fellow}
          }

   \offprints{S. Lopez, slopez@das.uchile.cl}

   \institute{ Departamento de
          Astronom\'{\i}a, Universidad de Chile, Casilla 36-D,
          Santiago, Chile.
\and
Hamburger Sternwarte, Universit\"at Hamburg, Gojenbergsweg 112,
21029 Hamburg, Germany.
\and
ESO, Karl-Schwarzschildstr. 1, Garching b. Muenchen, Germany.
\and
The Observatories of the Carnegie Institute of Washington, 813 Santa
Barbara St., Pasadena, CA 91101, USA.
}

   \date{}
   \authorrunning{S. Lopez et al.}
   \titlerunning{A dust-free damped \lya\ system at $z=2.3$}
   \abstract{
We have obtained a high resolution, high S/N UVES spectrum of the
   bright QSO \he\ to analyze the damped \lya\ system (DLA) observed
   at $z=2.33$. We measure column densities for \hone, \none, \altwo,
   \sitwo, \ptwo, \stwo, \arone, \crtwo, \fetwo, \nitwo, \zntwo,
   \ctwo\, \oone, and \althree; and put upper limits on the abundances
   of \mntwo, \sthree, \fethree, \pthree, and \ntwo. {\bf The metallicity of this
   system is $1/12$ solar at a neutral hydrogen column density of
   $\log N(\hone)=20.7$. From the observed ratios [Zn/Cr]
   $=-0.01\pm0.05$ and [S/Si] $=-0.06\pm0.03$ we conclude} that dust is
   {\bf very likely} absent from the ISM of this protogalaxy. We
   observe an enhancement of the $\alpha$/Fe-peak ratios of $+0.2$ dex
   for various elements, a marked odd-even effect in Mn, and a strong
   underabundance of N relative to Si and S, [N/Si,S] $=-1$ at [Si/H]
   $=-0.86$. All of these ratios support an environment that is in an
   early evolutionary stage, where the onset of star formation has
   begun only shortly before the DLA was observed.  We also perform a
   cloud-by-cloud analysis -- without precedent at high redshift --
   and find a tight correlation of all low-ionization species with
   respect to \fetwo\ extending over 2.5 orders of magnitude in
   $N(\fetwo)$. We interpret this trend as being due to homogeneous physical
   conditions (very mild ionization effects, common dust-destruction
   histories, same chemical composition) and propose that this line of
   sight encounters absorbing clouds that share a common
   environment. In addition, photoionization models show that these
   single clouds are shielded from the external ionizing radiation, so
   the fraction of ionized gas is small and, except for argon, does
   not influence the measured metal abundances. The {\bf observed}
   \althree/low-ion ratios suggest the mildly ionized gas occurs in
   shells surrounding neutral cores of \altwo.
      \keywords{Cosmology: observations --
                Quasars: individual: HE 2243$-$6031
                Quasars: general --
                absorption lines
               }
   }


   \maketitle

\section{Introduction}

Damped \lya\ systems (DLAs), defined as those QSO absorption systems
with the highest column densities, $N(\hone)> 2 \times 10^{20}$ \icm,
are widely believed to occur in the ISM of the high redshift
progenitors of present-day galaxies (Wolfe~\cite{Wolfe}). It is
fortunate that such large 
column densities permit very accurate metal abundance measurements
because, since DLAs contributed most of the neutral gas mass when the
Universe was 10\% of its present age (e.g., Storrie-Lombardi et
a.~\cite{Storrie}), they provide us with a powerful means to trace
the chemical evolution of star forming galaxies over a considerable
fraction of the Hubble time.  Since the absorbing gas in DLAs is expected to
be mostly neutral, the bulk of metals should be in neutral or singly
ionized form (depending on whether or not the first ionization
potential IP $>13.6$ eV). Thus, the many transitions of these ``low
ions'' that are redshifted to the optical range can be used to compute
[X$^+$/H$^0$] $\approx$ [X/H] $\equiv \log ({\rm X/H}) - \log ({\rm
X/H})_{\sun}$.

With the advent of 8-10m-class telescopes high resolution spectroscopy
of distant QSOs and detailed abundance studies in DLAs have
become possible  with relatively high accuracy ($\sim 10$ \%). 
The picture that has emerged is that of $Z\sim 1/10 -
1/300~Z_{\sun}$ absorbers with abundance patterns that, {\bf except for
the Fe-peak ratios},  resemble that
of Milky Way metal-poor halo stars (Lu et al.~\cite{Lu}; Pettini et
al.~\cite{Pettini}; Prochaska \& Wolfe~\cite{Prochaska}; Boisse et
al.~\cite{Boisse}). These observations seem to support the
underlying hypothesis that at least part of the DLAs have undergone a
similar chemical evolution history as our own Galaxy.

On the other hand, the observed sub-solar abundances [X/H] observed in
the Galactic ISM are interpreted as due to a partial incorporation of
the atoms from the gas phase into dust grains (e.g., Savage \&
Sembach~\cite{Savage}).  Since dust is formed through condensation from the
gas phase, any measure of the absorbing column density of a refractory
element will underestimate the true number of atoms. Moreover,
observations of diffuse clouds in the Galactic ISM have shown the
clear trend of different elements to condense into grains in different
proportions, depending on condensation temperature (Savage \&
Sembach~\cite{Savage}). Fe, Cr and Ni are among the most depleted
elements in the ISM, while O, P, Ar, S, and Zn have generally very
small depletion factors.  Since DLAs are likely to sample the
ISM of galaxies at earlier evolutionary stages, it seems reasonable to
raise the question as to whether the measured abundances at high redshift
have also been  modified by differential dust depletion (e.g., Lu et
al.~\cite{Lu}; Pettini et al.~\cite{Pettini}).  Comparing
abundances of refractory and non-refractory elements in high redshift
DLAs can lead to estimates of the dust-to-gas ratio in those
environments (Pettini et al.~\cite{Pettini}; Pettini et
al.~\cite{Pettini1}); assuming particular depletion patterns 
can in turn be used to correct the measured gas-phase
abundances (Vladilo et al.~\cite{Vladilo1}).  

However, any assessment of the
relative abundances in DLAs is  difficult as long as the interplay between
dust and nucleosynthesis effects is not well understood, and may not
be equal in high redshift DLAs and the local ISM.

The most direct way to disentangle these two contributions to the observed
abundance pattern is to find DLAs with as small a level of
depletion as possible. Such systems, though, are quite scarce and only
few cases are reported in the literature (Molaro et
al.~\cite{Molaro1}; Pettini et al.~\cite{Pettini2}; Prochaska \&
Wolfe~\cite{Prochaska3}).  {\bf In this paper we present VLT UVES
observations of 
the $z=3.01$ 
QSO \he\ (discovered by the
Hamburg/ESO QSO Survey; Reimers \& Wisotzki~\cite{Reimers1}), and argue
that the DLA observed at $z=2.33$ shows negligible amounts of
dust. 
We have been able to accurately measure
abundances for a set of elements of different nucleosynthetic origins
and with differing susceptibility to incorporation into dust
grains, including the detection of the ``rare'' elements P and Ar.}

The paper is organized in three parts. First we address the
question of the physical conditions in the {\it single} velocity components
and perform a cloud-by-cloud analysis of the relative abundances
(\S~\ref{variations}); next we look at the  uncertainties introduced by
possible ionization effects (\S~\ref{ionization}); and the final part
(\S~\ref{abundances}) is devoted to studying dust-depletion effects and
relative abundances. The conclusions are outlined in
\S~\ref{conclusions}. 

\section{Observations and  data reduction}

   \begin{table*}
      \caption[]{UVES Observations of \he}
         \label{tbl-1}
      \[
         \begin{tabular}{lccr}
            \hline
            \noalign{\smallskip}
{\rm Mode}&{\rm Wavelength}&{\rm Exp. Time}&{\rm Observing Date}\\
              &[nm]           &  [sec]        &                    \\
            \noalign{\smallskip}
            \hline
            \noalign{\smallskip}
Dichroic (346+580)    &308-388,477-675 &14\,400         &August 4, 5 and
      6, 2000\\ 
Dichroic (437+860)    &376-500,667-1040&10\,800        &August 8,11
      and 12, 2000\\
Red arm (520)         &477-676&3\,600                  &July 10, 1999\\
            \noalign{\smallskip}
            \hline
         \end{tabular}
      \]

   \end{table*}

\he\ was observed in service mode with the UVES instrument at the ESO Kueyen
telescope in August 2000 under good
seeing conditions ($\la 0.8\arcsec$). An extra 1-hour exposure had been
previously taken during
science verification of the instrument. Using two dichroic modes
the whole optical range was covered with a total exposure time
of $28\,800$ sec (Table~\ref{tbl-1}). 

After bias-subtracting and flat-fielding of the individual CCD frames,
the echelle orders were extracted and reduced with self-implemented
routines running under MIDAS. The algorithm attempts to reduce the
statistical noise 
to a minimum by fitting the seeing profile with Gaussians.  Flux
values are assigned with a variance according to the Poisson
statistics and the read-out noise, while cosmic-ray hits are 
assigned with infinite variances. The extraction proceeds in 3 steps: (1) 
A Gaussian with unconstrained parameters is  fitted to the sky profile
at each wavelength using the 
Levenberg-Marquardt method (Press et al. \cite{Press}); (2) 
the variation of width and position with respect to the orders 
is then fitted along the dispersion direction with low-order
polynomials; (3) step (1) is repeated, this time with width and
position fixed at the values given by the fit solution, so only the
amplitudes can vary.

The extracted orders were wavelength calibrated using as reference
 Th-Ar spectra taken after each science exposure.  All wavelength
 solutions were accurate to better than typically $1/10$ pixel. The
 wavelength values were converted to vacuum heliocentric values and
 each order of a given instrumental configuration was binned onto a
 common linear wavelength scale of $0.043$ \AA~pixel$^{-1}$. The
 reduced orders were then added with a weight according to the inverse
 of the flux 
 variances. Finally, the flux values  were normalized by a
 continuum that was defined using cubic splines over featureless
 spectral regions. The spectral resolution is FWHM $\sim 6.5$ \kms. The
 typical signal-to-noise ratio varies between S/N$\ga 80$ at
 3700--5700 \AA, and S/N$\ga 50$ at 5700--8000 \AA\ (becoming worse at
 both ends of the spectrum). 

{\bf In addition to the UVES data, a low resolution (FWHM $\approx 4$ \AA)
spectrum of \he\ was obtained on October 12 2001 using the Boller and
Chivens (B\&C) Spectrograph on the Baade 6.5m Telescope at Las Campanas
Observatory. This spectrum was used to better define  the quasar continuum
in the spectral region around the damped \lya\ line. }

\section{Line parameters}

Column densities $N$ and Doppler parameters $b$ were obtained by
fitting Voigt profiles to each velocity component. 
The fits were performed using the minimum-$\chi^2$ code FITLYMAN
(Fontana \& Ballester~\cite{Fontana}). 
Since most of the species under study present 2
or more transitions, the results of the fits were in general 
stable and quite independent of the fitting region chosen. 

As an alternative approach to obtain column densities in DLAs, the
apparent optical depth method (Savage \& Sembach~\cite{Savage1}) has
been often used to get integrated column densities. The method is
straightforward and gives {\bf accurate results for a resolved and
non-saturated line} (Lu et al.~\cite{Lu}; Prochaska \& 
Wolfe~\cite{Prochaska}).  In our case, although most of the ions have
at least one transition likely not to be saturated, we use instead the
fitting approach, because it provides us with better column density
estimates of partially blended components, and of those lines having
evident departures from the linear part of the curve-of-growth. In
addition, unlike previous studies, in which redshift and Doppler
parameter of a given velocity component are held fixed for all ions,
in our fitting procedure all fit parameters were allowed to
vary.\footnote{{\bf Traditionally, line profile fitting of DLA systems has 
been performed} by assuming the same Doppler parameter for all ions at
given velocity component, i.e., that bulk motion dominates over
thermal motion. This is a reasonable choice as long as DLAs occur in
cold gas and the ions under study are massive; however, this does not
prevent us {\it a priori} to chose the free-parameter approach
described above.}

The fit results are summarized in Table~\ref{tbl-2}, while the column
labeled 
$N_{\rm fit}$ in Table ~\ref{tbl-5} lists the overall column densities
which resulted from summing the individual component column densities.
All line parameters come from the Voigt profile
fits, with the exception of $N(\zntwo)$ and $N(\crtwo)$, which were
obtained using the apparent optical depth 
method. Following is  a
description of the fits to individual species.

\subsection{Description of the fits}
\label{fits}

%
   \begin{figure}
      \vspace{0cm}
\hspace{0cm}\epsfig{figure=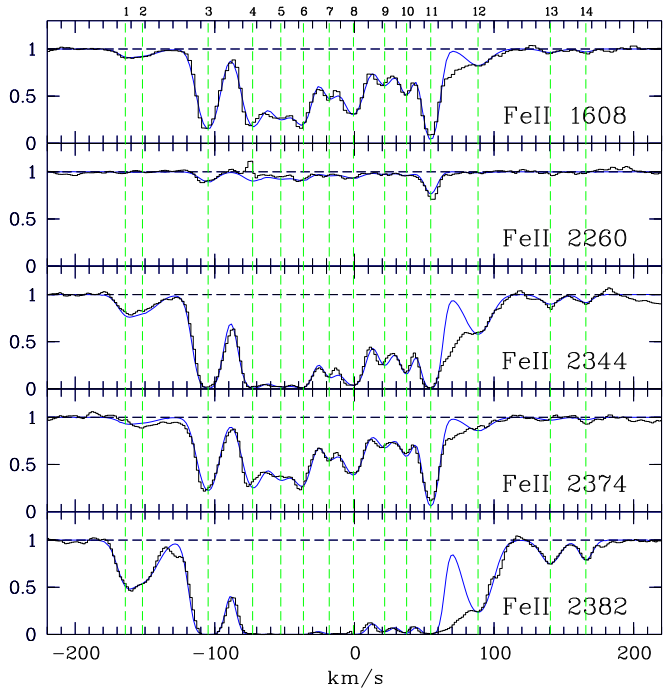,width=8.8cm}
\vspace{0cm} \caption[]{ 
Velocity plots of observed \fetwo\ transitions (histogram) and fitted
profiles. The dashed lines indicate the positions of the fitted
velocity components. The zero-velocity point corresponds to $z=2.330000$.

 } \label{fig_FeII}
      \end{figure}

\paragraph{\fetwo.---}

We begin by fitting \fetwo\ because this is the ion with the most complete set
of transitions redward of the \lya\ forest.  We fitted simultaneously
5 \fetwo\ transitions. They are shown in Fig.~\ref{fig_FeII}, plotted
in velocity space with respect to $z=2.330000$, and along with the fitted
profiles (smooth line). {\bf The $\lambda 2344$ and $\lambda 2382$
transitions 
provided a fitting interval only at the position of the weaker
components.} We found that 14 velocity components were necessary to
model the whole profiles. Their positions are indicated by the
vertical dashed lines. We used these line positions to identify -- but
not to tie -- 
velocity components in the remaining ions and throughout the paper we
refer to them as to ``C1'', ``C2'', etc. We were not able to fit the
absorption feature between C11 and C12. There appears not to be enough
information on the line profile for FITLYMAN to find the best line
centroid. Instead of defining a fixed line position `by hand' we
preferred to leave the feature unfitted. An optical depth analysis
shows that the contribution to the total \fetwo\ column density should
be less than 1\%.

%
   \begin{figure}
      \vspace{0cm}
\hspace{0cm}\epsfig{figure=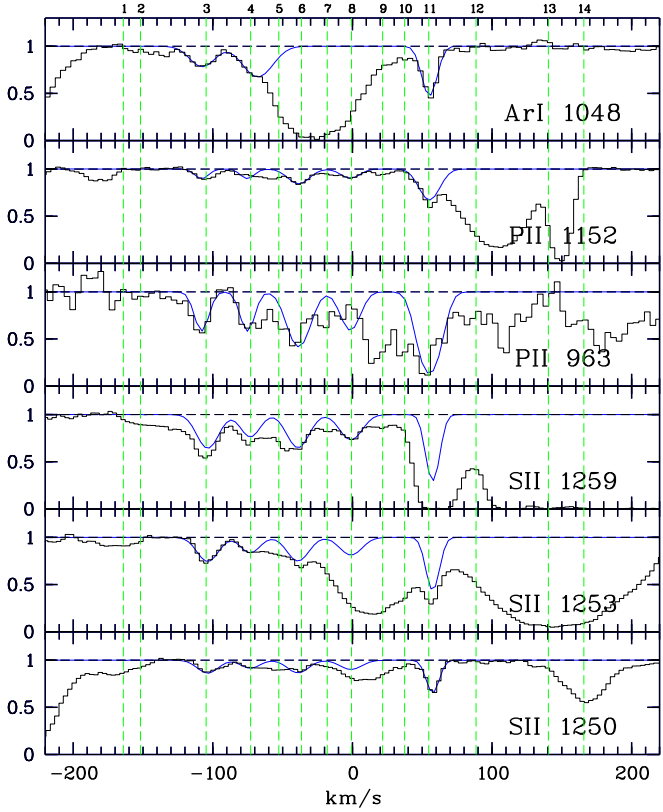,width=8.8cm}
\vspace{0cm} \caption[]{ 
Same as in Fig.~\ref{fig_FeII} but for the observed \arone, \ptwo\ and
\stwo\ transitions. The dashed lines correspond to the \fetwo\
components. 
 } \label{fig_ArI}
      \end{figure}

\paragraph{\arone.---}

We have detected \arone\ via the $\lambda 1048$ transition. The
$\lambda 1066$ line unfortunately does not provide any useful
constraint to the fits.  Although the $\lambda 1048$ transition lies
in the forest, the spectrum is clean at the expected position of its
strongest components. Absorption by \arone\ is detected in C3, C4 and
C11, which correspond to the three strongest lines in \fetwo\ ($\sim
50$ \% of the total \fetwo).  {\bf The small line widths suggest these
lines are very likely not} due to weak \lya\ interlopers. Moreover, the
position of the lines match well those of \fetwo, with exception maybe
of C4, due to blending with a \lya\ forest line (which makes the fit
an overestimate of $N(\arone)$ in this component). This is the second
ever measurement of argon at high redshift after the $z=3.39$
detection by Molaro et al. (\cite{Molaro}), and one of the few outside
the Galaxy available (Vidal-Madjar et al.~\cite{Vidal-Madjar};
Levshakov, Kegel \& Agafonova~\cite{Levshakov}).

\paragraph{\ptwo.---}

We have detected \ptwo\ via the $\lambda 963$ and $\lambda1152$
transitions (Fig.~\ref{fig_ArI}). 
We are quite confident of the detection
and fit results because they are based on two transitions that happen
to fall in a relatively unabsorbed part of the forest. 
This is only the third detection of extragalactic
phosphorus so far (Molaro et al.~\cite{Molaro}, Outram et
al.~\cite{Outram}). The fits give reasonably good results for C3, C4,
C6, and C8. 
{\bf C11 is blended with a \lyb\ line at $z=2.743903$ with $\log N =
14.57$ and $b=25.1$. These parameters resulted from fitting
simultaneously the \ptwo\
lines, the \lyb, and its corresponding \lya.}

\paragraph{\stwo, \none, \sitwo, \nitwo, \altwo.---}

We did not encounter big difficulties by fitting these ions because
all of them but \altwo\ show more than one transition, although we
were able to fit all 14 components only for \sitwo. The \altwo\
$\lambda 1670$ line is saturated over most of the components, so 
column densities of the stronger components must be treated as lower
limits.  In \stwo\ and \none\ there is obvious blending with forest
lines. C1 and C2 in \none\ were treated as one single component.

\paragraph{\oone, \ctwo.---}

For these ions most transitions have saturated components or they lie in
the forest, so only a small set of components could be fitted.  C3 in
\ctwo\ and C8 in \oone\ are uncertain because of the small fitting
interval available. C1 and C2 in \oone\ were treated as one single
component. 

\paragraph{\mntwo.---}

\mntwo\ is not detected (only the weak $\lambda 2606$ is available,
the other transitions falling in the red cross-disperser gap at $\sim
8600$ \AA). We derive a $3\sigma$ upper limit of $\log N<11.6$ for a
single line. If this value is scaled according to the strongest
\fetwo\ component, it gives a total \mntwo\ $3\sigma$ limit of $\log
N(\mntwo)<12.43$. 

\paragraph{\zntwo\ and \crtwo.---}
\label{zinc}


To estimate column densities for \zntwo\ and \crtwo\ 
we used the \zntwo\ $\lambda 2026$ and \crtwo\ $\lambda 2056$
transitions. These are the strongest transitions  available and should not be
blended with unrelated lines. The $\lambda 2062$ line is partially
lost within telluric lines. 

Unfortunately, the $\lambda 2026$ and $\lambda 2056$ lines lie in
the noisiest part of our spectrum, where spurious features introduced
by traces of fringing could not be removed properly. In addition, the
\zntwo\ line happens to fall in the overlapping region of two echelle
orders, where the continuum definition is inaccurate.  This was
reflected in the impossibility of profile fitting all components.
For this reason we used the apparent optical depth method to constrain
the respective column densities. {\bf Integrating between $v=-180$ and
$v=+100$ \kms\ leads to 
$\log N(\zntwo) = 12.22\pm 0.03$ and $\log N(\crtwo) = 13.24\pm
0.02$. We have excluded the velocity region [+100,+180],} 
due to spurious features that introduce extra
noise (excluding C13 and C14 introduces an uncertainty no larger than
$\sim 0.5$\%, though). We stress, however, that due to the weakness of
the lines  
these values are subject to continuum uncertainties (which are not
considered in the formal errors).

%
   \begin{figure}
      \vspace{0cm}
\hspace{0cm}\epsfig{figure=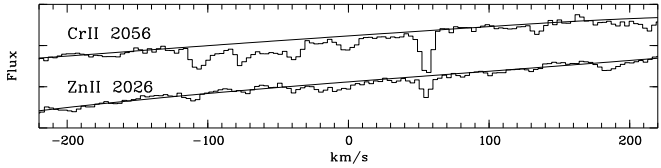,height=8.8cm,angle=-90}
\vspace{0cm} \caption[]{ 
Portion of the echelle order containing the \crtwo\  $\lambda 2056$ line and
portion of one of the orders containing the \zntwo\ $\lambda
2026$. The flux scale was accomodated for displaying purposes, and the
zero-velocity point corresponds to $z=2.330000$.

 } \label{fig_zinc}
      \end{figure}

   \begin{figure}
      \vspace{0cm}
\hspace{0cm}\epsfig{figure=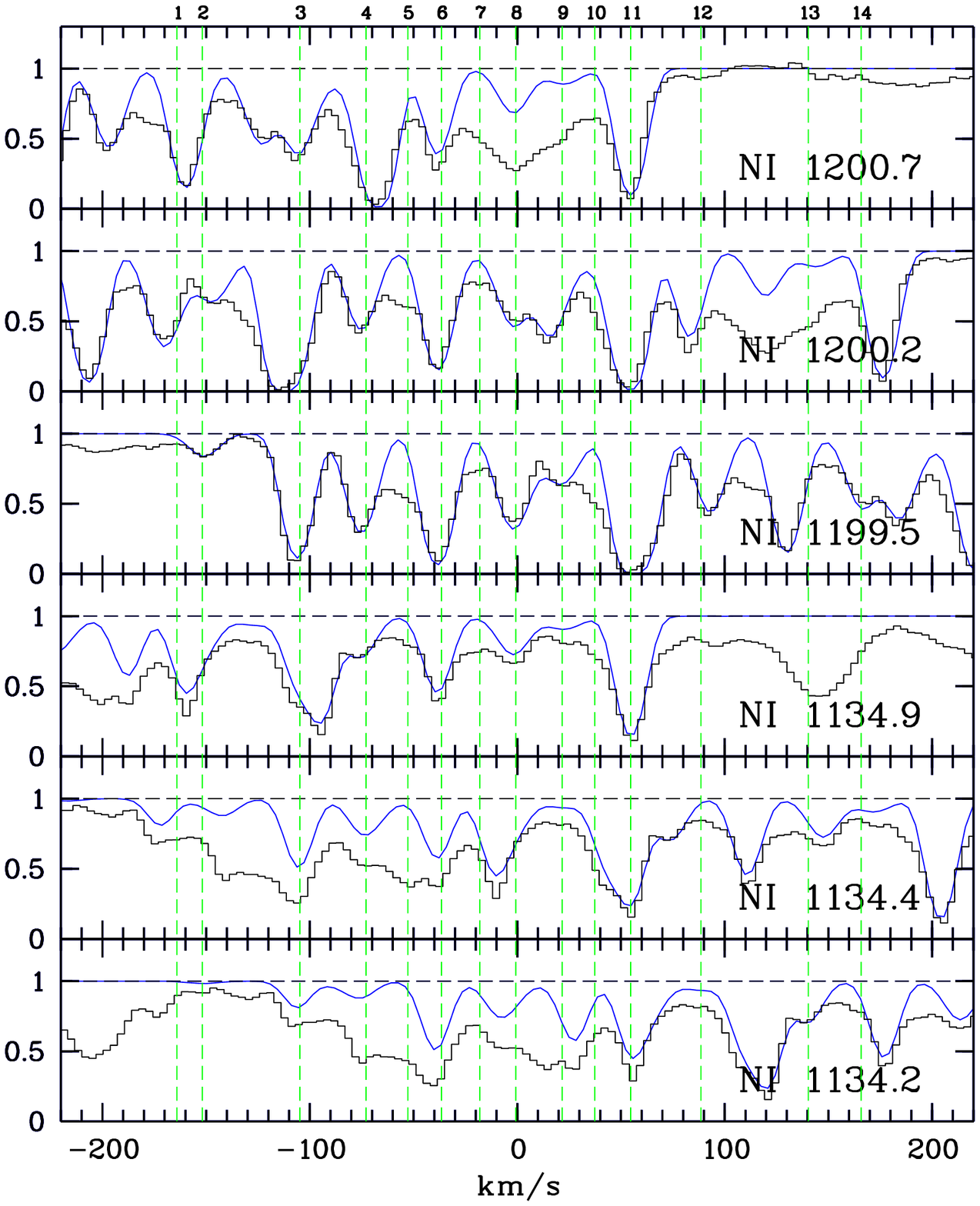,width=8.8cm}
\vspace{0cm} \caption[]{ 
Same as in Fig.~\ref{fig_ArI} but for the observed \none\ transitions.

 } \label{fig_NI}
      \end{figure}
%
   \begin{figure}
      \vspace{0cm}
\hspace{0cm}\epsfig{figure=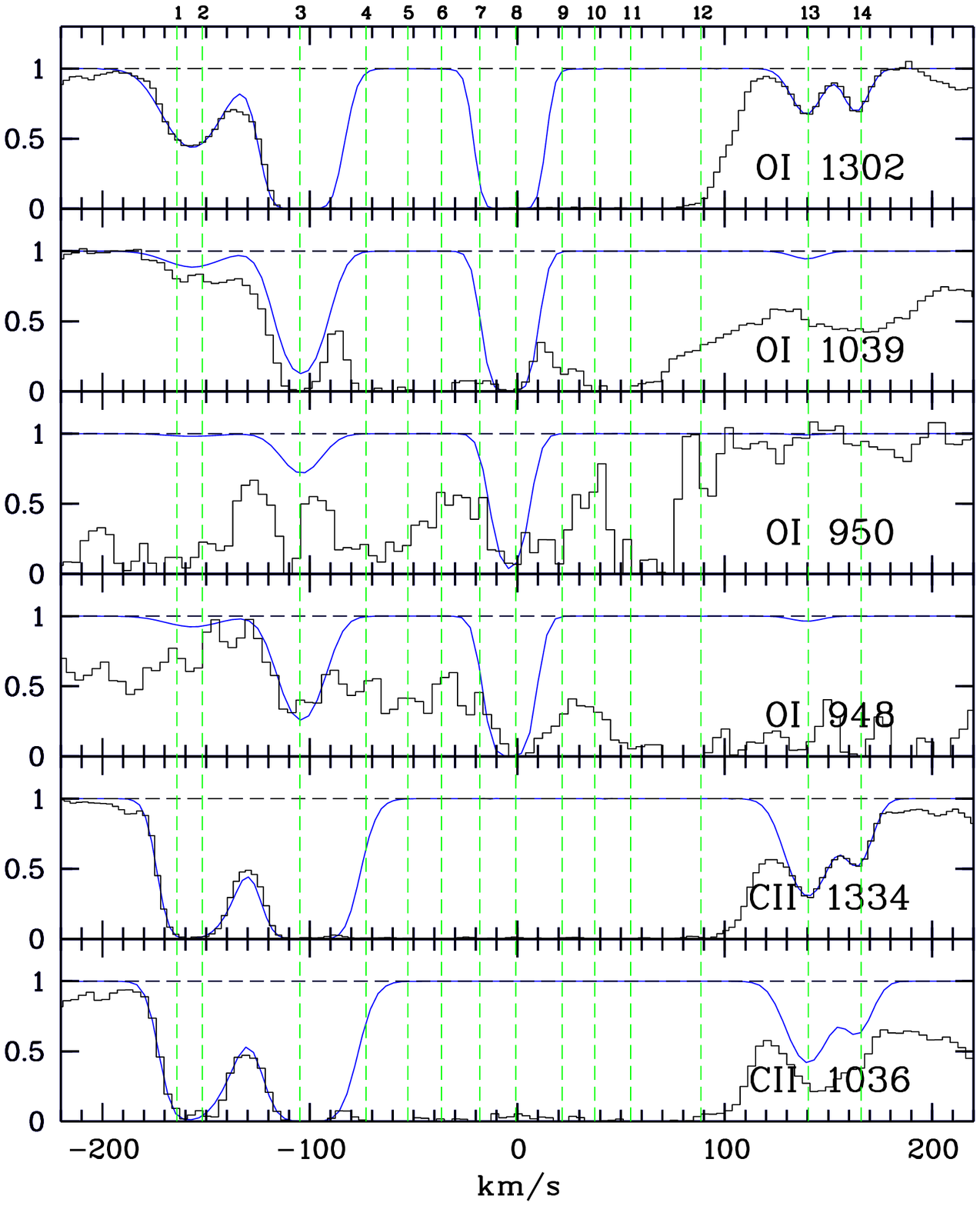,width=8.8cm}
\vspace{0cm} \caption[]{ 
Same as in Fig.~\ref{fig_ArI} but for the observed \oone\ and \ctwo\
transitions. 
 } \label{fig_OI}
      \end{figure}
%
   \begin{figure}
      \vspace{0cm}
\hspace{0cm}\epsfig{figure=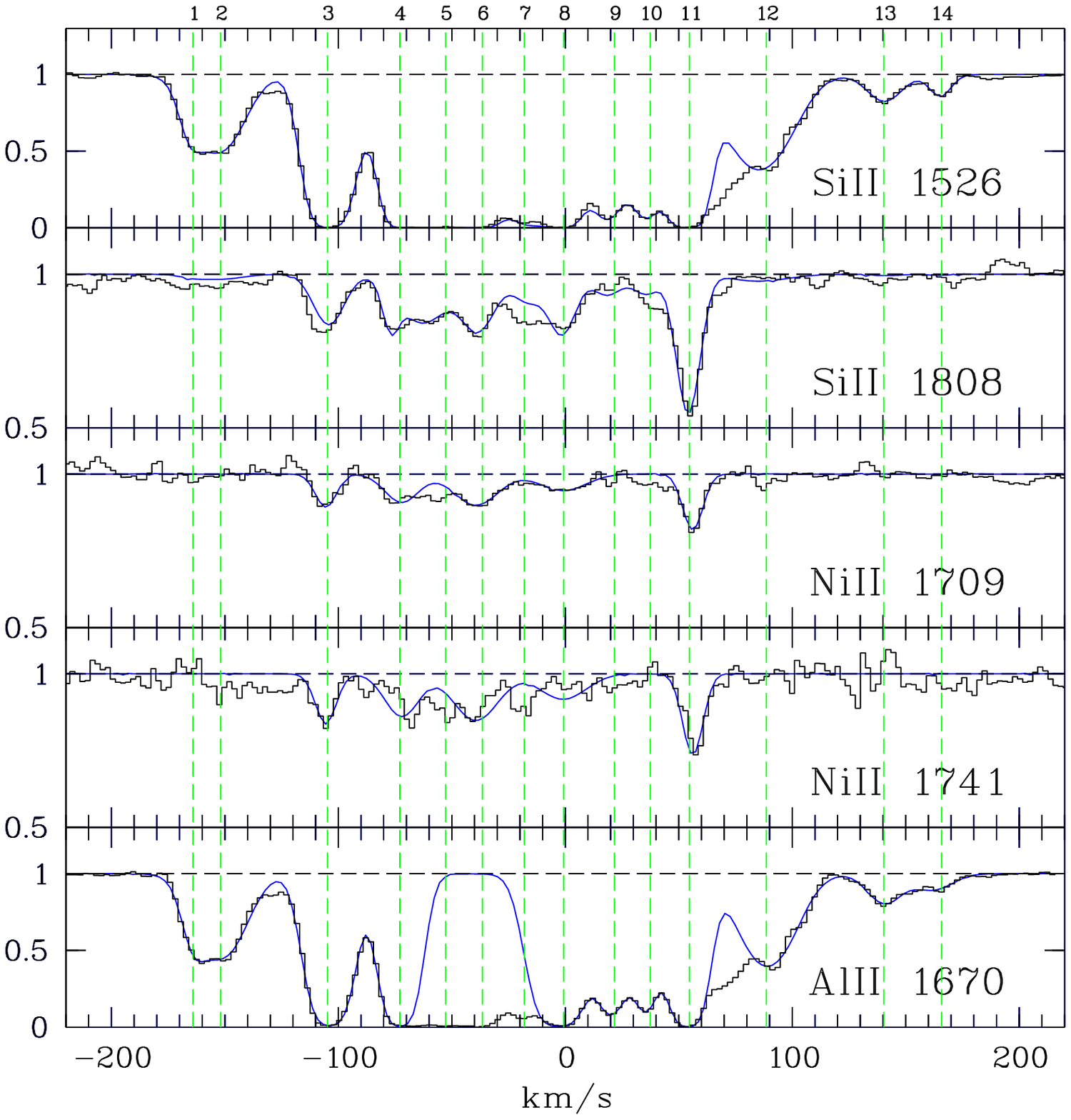,width=8.8cm}
\vspace{0cm} \caption[]{ 
Same as in Fig.~\ref{fig_ArI} but for the observed \sitwo, \nitwo\ and \altwo\
transitions. 

 } \label{fig_AlII}
      \end{figure}

\paragraph{Doubly ionized species.---}

\begin{figure}
      \vspace{0cm}
\hspace{0cm}\epsfig{figure=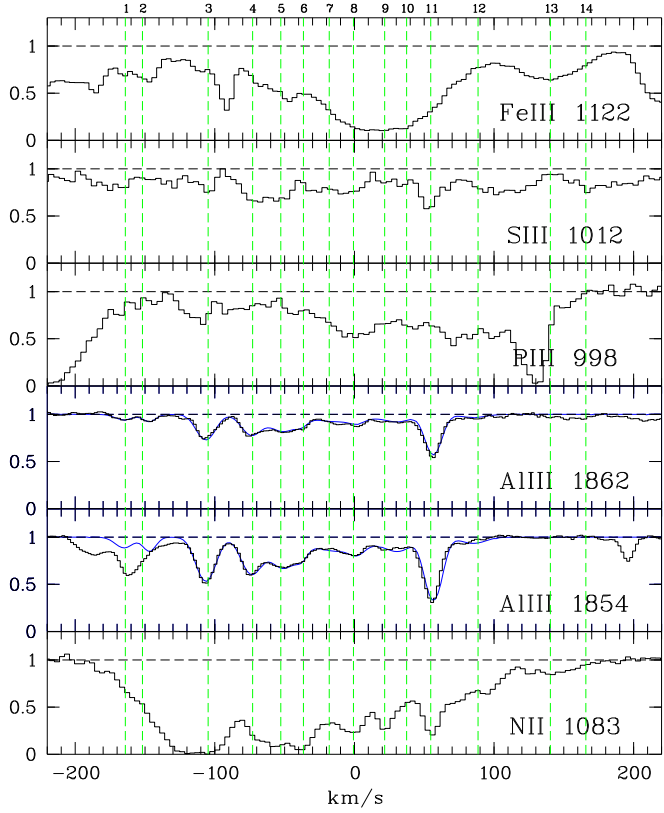,width=8.8cm}
\vspace{0cm} \caption[]{ 
Same as in Fig.~\ref{fig_ArI} but for 
doubly ionized species and \ntwo. The absorption
feature at $v=-90$ \kms\ in the top plot is unrelated from \fethree\
$1122$. 

 } \label{fig_medium}
      \end{figure}

Fig.~\ref{fig_medium} displays the expected position in velocity space
of all doubly ionized species covered by our spectrum. {\bf \ntwo\ is also
included in the figure.} It is of great
importance to estimate or put useful limits on column densities
for these ions, as they will 
constrain the ionization level of the system.  

For \althree\ we have at our disposal two transitions with apparently
unsaturated lines, so the fit results can be considered reliable. Note
that C13 and C14 are not detected. 

For the rest of the ions we have
attempted to put upper limits on column densities using different
approaches. 
For \sthree\ we use the transition at $\lambda = 1012$ \AA. Although
the line lies in
the forest, there is a relatively good match of the absorption profile with
that of \fetwo, in line position as well as in line strengths. This
suggests the \sthree\ is present in the DLA 
gas and tracks the low-ion profiles. However, since the 
lines stand out only poorly we did not attempt a fit; instead, we
used  the total apparent column density, which gives 
$N(\sthree)=14.05\pm 0.01$ for $v=$ [$-130,+80$]. 

To constrain \fethree\ we use the $\lambda 1122$ transition. This line
is also in the forest and appears so severly contaminated that no
resemblance with \althree\ or low-ion profiles is recognized.  An upper
limit was estimated by assuming a \fethree\ line lies at C3 and has
$b=b_{\fetwo}=8.0$ \kms. The maximum column density allowed by the
data then becomes $\log N=13.4$, which, weighted with C3 from the
\althree\ fits yields a total column density of $N(\fethree)<14.18$.
Note, however, that the \fethree\ profile does not exclude
$N(\fethree) = 0$. 

For \pthree\ we performed the same exercise using the expected position
of the $\lambda 998$ transition, which also shows no black
absorption. At C3 we get $\log N(\pthree)<13.2$
for $b=8.4$ \kms, or $\log N(\pthree)<13.98$ \icm\ for the whole
profile. Again, since the line is in the forest and we do not
recognize an \althree-like absorption pattern, we cannot
exclude $N(\pthree)=0$.

{\bf Finally, the $\lambda 1083$ 
profile, despite forest contamination,  also resembles the low-ion
pattern. This is interesting given  that nitrogen is expected to be
mainly neutral due to its high IP of $14.5$ eV. Assuming that the
absorption feature at C11 is a \ntwo\ $\lambda 1083$ line and
weighting with C11 from the \althree\ fits we estimate that 
$N(\ntwo)<14.55$. }

%

\subsection{Neutral hydrogen column density}

\begin{figure}
      \vspace{0cm}
\hspace{0cm}\epsfig{figure=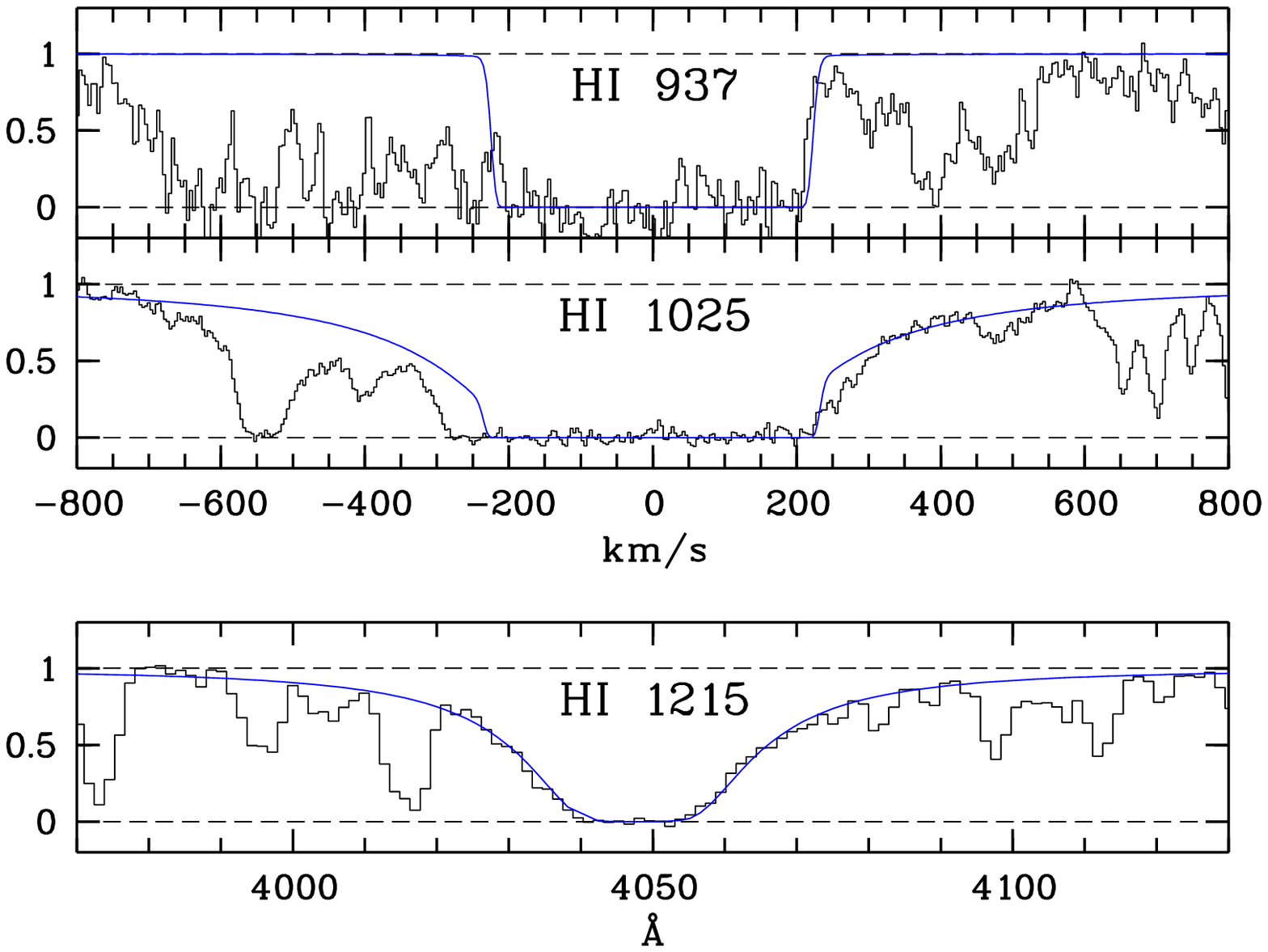,width=8.8cm}
\vspace{0cm} \caption[]{ 
{\bf 
UVES (top panels) and
B\&C spectrum of \he\ at the position of  the \hone\ lines used to constrain
$N(\hone)$. The smoothed line is the superposition of $14$
\hone\ velocity components centered at the positions of the \fetwo\ lines.
The individual \hone\ column densities are proportional to the \fetwo\
column densities, and the total \hone\ column density $\log
N(\hone)=20.67$ \icm. The Doppler parameter $b$ is assumed to be the same for
all components and has the value of $b=16$ \kms, as constrained by 
\lyb\ and Ly$\epsilon$.

}

 } \label{fig_DLA}
      \end{figure}

{\bf 
Despite its high S/N and coverage of higher order Lyman series lines,
obtaining $N(\hone)$ from the UVES spectrum alone posed some
difficulties owing to the non-Voigt shape of the damped \lya\ line. To
estimate $N(\hone)$ we used instead the B\&C spectrum, where the fit
to \lya\ was much better constrained than in the high resolution
spectrum. This shows the difficulties of defining a continuum around
an absorption feature that extends over several echelle orders.

We fitted the damped \lya\ line with 14 individual components whose
positions were fixed at the  \fetwo\ velocity
components. The \hone\ column density was varied as to keep the
$N(\fetwo)/N(\hone)$ ratio constant for all components, and the
$b$-parameter was assumed the same for all components. The best fit
we obtained implies an overall column density of $\log N(\hone)=
20.67\pm 0.02$ \icm. Since the same  
value is obtained if a single fit component is assumed, we conclude
that the assumption of constant \fetwo/\hone\ is unimportant in so far
as the $N(\hone)$ value is concerned. The fitted profiles are shown in 
Fig.~\ref{fig_DLA}.

We note that
the fit solution still leaves underabsorbed ``voids'' in the blue wings
of the \lyb\ line. These are produced by low column-density \lya\
systems that do not show
the low-ionization species studied in this paper.
}

Of course the underlying assumption in
our fitting procedure that \fetwo/\hone\ remains constant over all
velocity components, although expected if the gas is well mixed, must
not hold if \fetwo\ is for instance inhomogeneously dust
depleted. Next section deals with such possible variations.
%

   \begin{table*}
      \caption[]{Line parameters}
         \label{tbl-2}
      \[
\setlength{\extrarowheight}{2pt}
         \begin{tabular}{lrccr|lrccr}
            \hline
            \hline
{\rm Ion}&Comp.&$z$&$\log N(\sigma_{\log N})$&$b(\sigma_b)$ &{\rm Ion}&Comp.&$z$&$\log N(\sigma_{\log N})$&$b(\sigma_b)$ \\
         &     &   &\icm                     &\kms          &         &     &   &\icm                     &\kms          \\
            \hline

\ctwo  &    1  & 2.328224  &   ~~14.35    0.02$^a$&  9.4  0.2  &\stwo   &  3  &2.328849  &  14.17    0.01  & 9.5  0.5 \\
       &    2  & 2.328370  &   ~~14.04    0.01$^a$& 12.9  0.2  &        &  4  &2.329187  &  13.91    0.07  & 8.4  1.9 \\
       &    3  & 2.328879  &   ~~14.81    0.10$^a$& 15.5  1.8  &        &  6  &2.329567  &  14.17    0.03  & 9.7  0.9 \\
       &   13  & 2.331557  &   13.76    0.01  & 11.9  0.4  &        &  8  &2.329989  &  14.04    0.04  &10.1  1.4 \\
       &   14  & 2.331823  &   13.34    0.02  &  8.0  0.4  &        & 11  &2.330639  &  14.43    0.01  & 4.9  0.3 \\
\none  &  1+2  & 2.328322  &   12.87    0.04  &  9.2  1.2  &\arone  &  3  &2.328805  &  12.89    0.04  &11.5  1.4 \\
       &    3  & 2.328823  &   13.90    0.01  &  7.3  0.1  &        &  4  &2.329247  &  13.17    0.13  &13.6  4.2 \\
       &    4  & 2.329165  &   13.68    0.01  &  8.3  0.2  &        & 11  &2.330613  &  13.17    0.02  & 5.7  0.5 \\
       &    6  & 2.329573  &   14.01    0.00  &  7.5  0.1  &\fetwo  &  1  &2.328177  &  12.51   0.016  & 6.8  0.0 \\
       &    8  & 2.329980  &   13.68    0.01  &  9.1  0.3  &        &  2  &2.328314  &  12.86   0.008  &12.8  0.0 \\
       &    9  & 2.330252  &   13.21    0.03  & 10.3  1.2  &        &  3  &2.328836  &  14.09   0.003  & 8.0  0.0 \\
       &   11  & 2.330590  &   14.40    0.01  &  7.0  0.1  &        &  4  &2.329190  &  14.03   0.003  & 7.9  0.0 \\
\oone  &  1+2  & 2.328260  &   14.18    0.01  & 16.9  0.4  &        &  5  &2.329414  &  14.04   0.003  &10.6  0.0 \\
       &    3  & 2.328844  &   15.28    0.04  & 12.0  0.2  &        &  6  &2.329595  &  13.92   0.004  & 6.9  0.0 \\
       &    8  & 2.329960  &   ~~16.15    0.30$^a$&  7.4  1.2  &        &  7  &2.329800  &  13.60   0.005  & 6.9  0.0 \\
       &   13  & 2.331549  &   13.60    0.01  &  8.9  0.4  &        &  8  &2.329993  &  13.91   0.003  & 8.7  0.0 \\
       &   14  & 2.331820  &   13.45    0.01  &  6.0  0.3  &        &  9  &2.330241  &  13.45   0.005  & 7.1  0.0 \\
\altwo &    1  & 2.328179  &   11.76    0.03  &  5.1  0.3  &        & 10  &2.330415  &  13.52   0.005  & 5.7  0.0 \\
       &    2  & 2.328313  &   12.40    0.01  & 13.5  0.2  &        & 11  &2.330608  &  14.31   0.006  & 4.8  0.0 \\
       &    3  & 2.328846  &   ~~13.03    0.01$^a$&  8.8  0.1  &        & 12  &2.330984  &  13.20   0.005  &11.0  0.0 \\
       &    4  & 2.329190  &   ~~13.07    0.06$^a$&  7.2  0.3  &        & 13  &2.331559  &  12.39   0.018  & 8.2  0.5 \\
       &    8  & 2.329967  &   ~~13.14    0.02$^a$& 10.5  0.5  &        & 14  &2.331842  &  12.17   0.025  & 5.2  0.6 \\
       &    9  & 2.330226  &   12.64    0.02  &  6.8  0.4  &\nitwo  &  3  &2.328828  &  13.00   0.03   & 5.1  0.6 \\              
       &   10  & 2.330395  &   12.56    0.02  &  5.7  0.4  &        &  4  &2.329200  &  13.15   0.11   & 9.5  2.6 \\
       &   11  & 2.330601  &   ~~13.21    0.05$^a$&  6.3  0.3  &        &  6  &2.329562  &  13.30   0.06   &12.8  2.2 \\
       &   12  & 2.330989  &   12.50    0.01  & 15.1  0.3  &        &  8  &2.329989  &  13.04   0.10   &13.5  4.0 \\
       &   13  & 2.331562  &   11.69    0.04  &  9.4  0.8  &        & 11  &2.330624  &  13.24   0.02   & 4.7  0.4 \\
       &   14  & 2.331798  &   11.47    0.06  & 11.2  1.5  &\althree&  1  &2.328165  &  11.82   0.09   & 7.8  2.7 \\            
\sitwo &    1  & 2.328164  &   12.83    0.05  &  4.8  0.5  &        &  2  &2.328366  &  11.82   0.08   & 5.1  1.6 \\
       &    2  & 2.328302  &   13.55    0.01  & 13.8  0.3  &        &  3  &2.328815  &  12.57   0.01   & 8.4  0.2 \\
       &    3  & 2.328842  &   14.38    0.01  &  8.4  0.1  &        &  4  &2.329176  &  12.47   0.01   & 8.9  0.3 \\
       &    4  & 2.329145  &   14.04    0.06  &  2.9  0.8  &        &  5  &2.329428  &  12.50   0.02   &12.5  0.6 \\
       &    5  & 2.329301  &   14.56    0.02  & 13.9  0.5  &        &  6  &2.329579  &  11.67   0.11   & 4.2  1.3 \\
       &    6  & 2.329574  &   14.48    0.02  &  9.6  0.4  &        &  7  &2.329858  &  12.40   0.07   &24.3  4.3 \\
       &    7  & 2.329815  &   14.04    0.03  &  7.5  0.6  &        &  8  &2.330019  &  11.68   0.09   & 6.4  1.5 \\
       &    8  & 2.329983  &   14.42    0.03  &  7.2  0.3  &        &  9  &2.330250  &  11.64   0.19   &13.3  0.0 \\
       &    9  & 2.330204  &   13.94    0.05  &  7.6  1.3  &        & 10  &2.330363  &  11.98   0.04   &11.3  0.0 \\
       &   10  & 2.330397  &   13.89    0.08  &  7.0  1.9  &        & 11  &2.330626  &  12.75   0.01   & 6.9  0.1 \\
       &   11  & 2.330602  &   14.79    0.02  &  5.7  0.3  &        & 12  &2.330942  &  11.72   0.05   &11.7  1.2 \\
       &   12  & 2.330949  &   13.78    0.01  & 17.5  0.4  &        & 13  &2.331559  &$<$11.07         &          \\
       &   13  & 2.331562  &   12.84    0.02  &  9.6  0.5  &        & 14  &2.331842  &$<$11.07         &          \\
       &   14  & 2.331835  &   12.55    0.02  &  5.4  0.6  & &&&&\\                                                 
\ptwo  &    3  & 2.328801  &   12.26    0.08  &  5.3  1.3  & &&&&\\                                                 
       &    4  & 2.329165  &   12.27    0.14  &  5.3  3.4  & &&&&\\                                                 
       &    6  & 2.329572  &   12.65    0.07  &  9.6  2.2  & &&&&\\                                                 
       &    8  & 2.329983  &   12.37    0.11  &  8.3  3.2  & &&&&\\                                                 
       &   11  & 2.330672  &   12.97    0.02  &  8.4  0.4  & &&&&\\                                                 

\hline

         \end{tabular}
      \]

\begin{list}{}{}
\item[$^a$] {\small Based on saturated transitions}
\end{list}

   \end{table*}

\section{Inferences from cloud-to-cloud variations} 
\label{variations}

In this section we investigate cloud-to-cloud variations of the
relative abundances. There are 3 possible sources of abundance
variations among clouds: (1) differential depletion due to differing
composition and destruction histories  of dust grains; (2) different ionization
conditions due to density variations and/or local changes in the
ionization field (\S~\ref{ionization}); and (3) simply differing
chemical compositions.   

\subsection{No abundance gradients}

\begin{figure}
      \vspace{0cm}
\hspace{0cm}\epsfig{figure=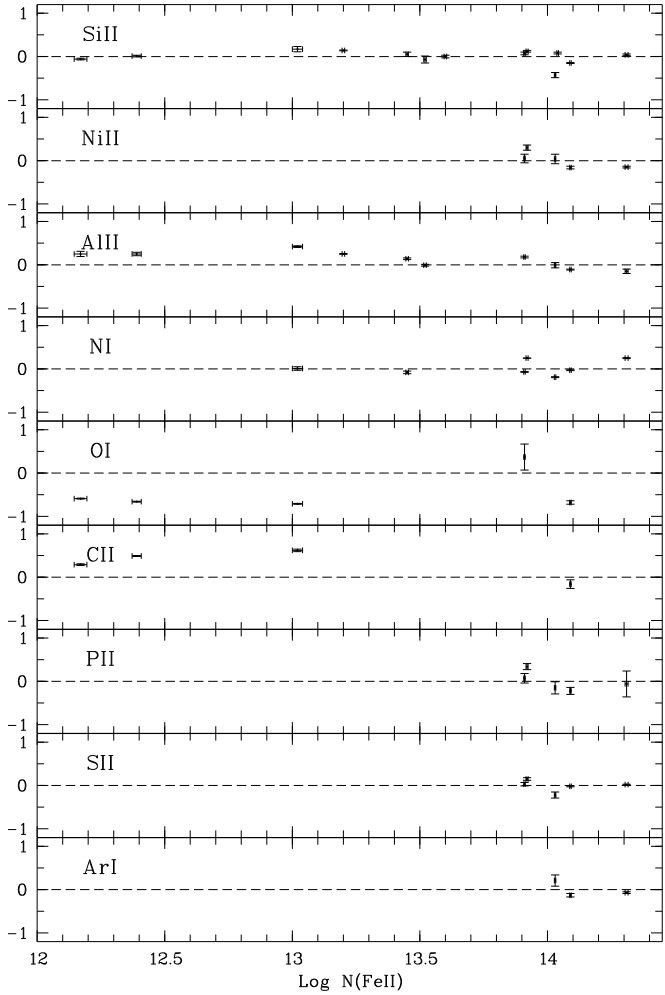,width=8.8cm}
\vspace{0cm} \caption[]{ 
{\bf $\log N_i({\rm X})/N_i(\fetwo) - \log N_{\rm corr}({\rm X})/N_{\rm
corr}(\fetwo)$ 
vs. $\log N_i(\fetwo)$ of all 
observed low-ion velocity components}. $N_{\rm
corr}$ denotes the total, corrected column densities listed in
Table~\ref{tbl-5}. 

 } \label{fig_variations}
      \end{figure}

{\bf Fig.~\ref{fig_variations} plots $\log N_i({\rm X})/N_i(\fetwo)$ versus
$\log N_i(\fetwo)$ for
all velocity components for which column densities and associated
errors are available. In each panel the values have been normalized by
$ N_{\rm corr}({\rm X})/N_{\rm corr}(\fetwo)$,}
where $N_{\rm corr}$ is the total column density listed for
each element  in Table~\ref{tbl-5}.
%
%
Since we were not able to obtain individual component values of 
$N(\zntwo)$ we use instead $N(\fetwo)$ as reference, bearing in mind
that iron is 
possibly subject to differential dust depletion.
C1 and C2 have been integrated and
treated as a single component due to possible blending that could have
affected the fits; the new value corresponds to the point at $\log
N(\fetwo)= 13.02$.  

%

From Fig.~\ref{fig_variations} we see that the $N({\rm X})/N(\fetwo)$
ratio remains, 
within measurement errors, fairly constant over almost two and half orders of
magnitude in $N(\fetwo)$. This is specially evident in \sitwo, for
which we have the most robust column density estimates and coverage of
all \fetwo\ components. The same trend holds even for those ions for which
we were able to obtain $N$ only for the strongest components (\nitwo, \none,
\ptwo, \stwo, \arone). The major discrepancies with respect to the
assumed abundances are found in \oone\ and \ctwo, but these are easily
explained by the saturated C3 in both transitions used for \ctwo, and
by the overfitted C8 in \oone. \altwo\ also shows discrepant values
between weak and stronger components; this is most likely  due to saturation
effects affecting the fits to high column densities components (which
underestimate the true  $N(\altwo)$).  

From the data on \sitwo\ we deduce that the single column densities
diverge from the assumed value $\log N(\sitwo)=\log N(\fetwo)+0.44$ on
average by only $0.15$ dex. 
The constancy of $N(\sitwo)$ over such wide range of \fetwo\ column
densities has important implications for this DLA. 
First, silicon is an $\alpha$-element. If the delayed injection of
processed material from SN Type I and II had undergone differing histories from
cloud to cloud, one would expect such variations to be reflected in
the individual \sitwo/\fetwo\
ratios; indeed, we would expect to observe as large a  scatter in
\sitwo/\fetwo\ as the dispersion observed among
different DLAs of the same metallicity (e.g., Lu et al.~\cite{Lu})). 
Secondly,  silicon is a refractory element. Locally,
the depletion of Si and Fe differ by much larger factors than what is
observed here, over all ISM environments from the cool disk to the warm
halo (Savage \& Sembach~\cite{Savage}). In addition, since the
gas-phase abundances of refractory elements decrease with
$\log{\hone}$ (e.g., Wakker \& Mathis~\cite{Wakker})
-- and thus with $\log N(\fetwo)$ in this DLA-- we would expect
departures from the 
assumed abundances at higher values of $\log N(\fetwo)$. Altogether,
the small variations 
we observe in Fig.~\ref{fig_variations} mean both the chemical evolution
in single clouds has been the same, and dust is either absent or has
different properties from the ISM. 
Such small variations
have been pointed out just for one other DLA (Prochaska \&
Wolfe~\cite{Prochaska2}) toward Q0201+365 at $z=2.5$. Those authors
also suggested the absence of dust could explain their observations. 
The novelty of our approach is that we present the first cloud-by-cloud
analysis where that trend is observed
(instead of integrating the column densities over particular velocity
windows).
From an instrumental point of view, the homogeneous
behavior of these abundances imply the $N(\fetwo)$-weighted
corrections that we will apply to some of the
low-ion column 
densities are 
physically acceptable.


The behavior we see in this DLA has not been observed in the local
ISM, which shows a range of values of roughly $1$ dex (Spitzer \&
Fitzpatrick~\cite{Spitzer}; Welty et al.~\cite{Welty1}), nor in the
LMC (Welty et al.~\cite{Welty1}). It is well known that such
lines of sight probe regions in different environments; therefore,  it seems
conceivable that this DLA encounters an homogeneous
population of clouds. 

Under the assumption that \fetwo\ tracks \hone\ this line of sight 
probes the interval $\log N(\hone)\approx 18 - 20$ \icm, which,
if in our Galaxy, would mean probing a mixture of halo and disk
clouds (Wakker \& Mathis~\cite{Wakker}; and references
therein). Wakker \& Mathis (\cite{Wakker}) find a remarkable 
anticorrelation of the metal abundance with \hone\ in the local ISM, which
they interpret as the environmental dependence expected if both disk
(with low gas-phase abundances) and halo (with high gas-phase
abundances) are plotted together as a function of \hone. Our observations do
not fit into this scenario, as -- again assuming $N(\hone) \propto
N(\fetwo)$ -- they show no dependence of
abundances with respect to \hone\ column density
(Fig.~\ref{fig_ionization_a}). Two possible explanations may be
responsible for this discrepancy: (1) \fetwo\ simply does not track
\hone\, but is an increasing function of \hone, in which  case, a higher
density cloud would have a higher metallicity; or (2) this particular line of
sight  probes just one type of environment. The above discussion seems
to support the second alternative.

\section{Ionization}
\label{ionization}

If high redshift DLAs occur in dense protogalactic clumps, then one
can expect a soft ionizing radiation field which is not able to
penetrate gas with $N(\hone)>10^{20}$ \icm. As a result, all elements
but those having their first ionization potential $>13.6$ eV:
\none, \oone, \arone, should be primarily in singly ionized form, with
little gas in ionized form (e.g., Viegas~\cite{Viegas}). More
recently, several authors have investigated the systematic effects of
ionized gas on the determination of metal abundances from the low-ionization
species (e.g., Howk \& Sembach~\cite{Howk}; Vladilo et
al.~\cite{Vladilo}).  

In the present DLA the moderately low overall $N(\hone)$, and the detection of some
doubly ionized ions such as \althree\ having similar
velocity profiles as their singly ionized counterparts, both suggest
ionization may play a role in bringing part of the low ions to
a higher, in principle undetectable  ionization level. 
In this section we argue against ionization
affecting considerably  our measurements of the metal
abundances even for the {\it single} clouds.

\subsection{Self-shielding}

\begin{figure}
      \vspace{0cm}
\hspace{0cm}\epsfig{figure=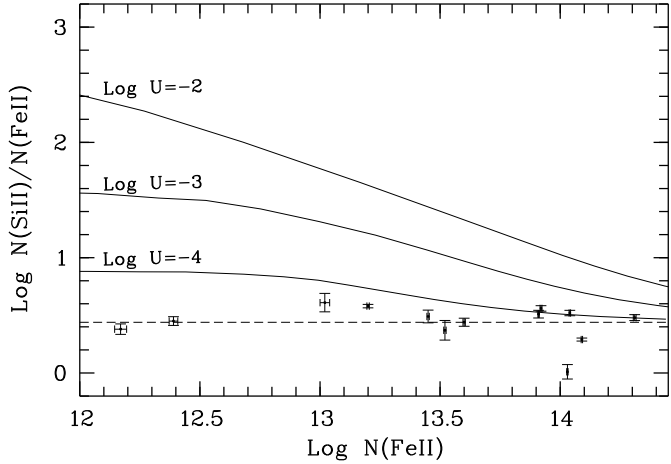,width=8.8cm}
\vspace{0cm} \caption[]{ 
Observed ratio $\log N(\sitwo)/N(\fetwo)$ vs. $\log N(\fetwo)$ along
with measurement errors. The dashed line mark the assumed ratio
from the overall \sitwo\ and \fetwo\ column densities. The solid
curves are CLOUDY simulations for an extragalactic ionizing field and
for [Fe/H] $=-1.0$ and [Si/Fe] $=+0.4$.

 } \label{fig_ionization_a}
      \end{figure}

Our observations suggest that \hone\ is distributed 
over all of the single components.  Under the assumption $\log
N(\hone)=\log N(\fetwo)+5.75$ for each of the velocity components, the
median value of an \hone\ cloud is derived to be $\log
N(\hone)=19.26$, i.e., at least half of the clouds would be
considerably ionized, were they part of a classical Lyman-limit system
(e.g., Lopez et al.~\cite{Lopez}). Given the complexity of the present
system, however, it is also conceivable that self-shielding does keep
the single clouds in a more neutral environment. Evidence for this
comes from the constancy of the low-ion/\fetwo\ ratio down to the
lowest measurable \fetwo\ column densities.
Fig.~\ref{fig_ionization_a} shows the ratio $\log
(N(\sitwo)/N(\fetwo))$ as a function of $\log N(\fetwo)$. 
The \sitwo/\fetwo\ ratio {\it of integrated column
densities} in DLAs should be only mildly influenced by ionization
effects (Vladilo et al.~\cite{Vladilo}; Howk \&
Sembach~\cite{Howk}). We interpret the datapoints in
Fig.~\ref{fig_ionization_a} as an extension of that property down to
much lower densities, thus suggesting self-shielding in the present
system.

Next, to address the question of how ionization may influence the
single-cloud  abundance ratios, we  compare the  fit results for the single
components of \fetwo\ and \althree, which are good proxies for the low
and medium-ionization gas, respectively. As for \fetwo, \althree\
column densities 
also fulfill the requirement of coming from unsaturated and resolved
lines. Ideally, we would have liked to have used \altwo\ instead of
\fetwo\ to disentangle 
the effect of possible variations in the relative abundances (despite
the results shown in \S~\ref{variations}), but the former ion has only
one transition ($\lambda 1670$), and some of the components are
saturated. Triply 
ionized species such as \cfour\ or \sifour\ were not used as their
profiles indicate they occur in distinct volumes. 
Fig.~\ref{fig_ionization_b} shows the ratio 
$\log (N(\fetwo)/N(\althree))$ as a function of $\log N(\fetwo)$ for
all 12 velocity components for which column densities and associated
errors are available.  The two lower limits correspond to the
non-detection of \althree\ at the positions of C13 and C14, and the
dashed line mark the \fetwo/\althree\ calculated from the overall
column densities.  We use the
same y-scale as in Fig.~\ref{fig_ionization_a}  to permit a comparison
of the scatters. 
Contrary to the constant behavior of the N(\sitwo)/N(\fetwo) ratio,
there is a larger scatter 
in $N(\fetwo)/N(\althree)$ of about 1 dex over the same interval in
$N(\fetwo)$. Given the constancy in the low-ion abundances, such scatter could
be attributed to cloud-to-cloud variations in ionization.

Our $\log N(\althree)/N(\altwo)\la-0.6$ is typical of DLAs with
similar $N(\hone)$, according to the compilation by Vladilo et
al. (\cite{Vladilo}). These authors find an empirical anti-correlation
between $N(\hone)$ and the \althree/\altwo\ ratio, suggesting that
\hone\ indirectly traces the level of ionization in the gas. A similar
trend is not seen in our \fetwo/\althree\ plot. We find that the two
points representing the most ionized clouds ({\bf i.e., lowest
\fetwo/\althree\ ratio; corresponding to C1+C2 and
C7)} contribute only 17 \% (6\%) of the total \althree\ (\fetwo)
column density. Ignoring those two datapoints we find that
\fetwo/\althree\ remains {\bf in this DLA} quite independent of
\fetwo\ (and thus of 
\hone\ in the single clouds). A linear regression analysis gives a
slope of only $m=0.16\pm 0.32$, which, should \fetwo\ indeed track
\hone\ over these interval, is small when compared with the $m=0.81\pm
\pm0.015$ found by Vladilo et al. (\cite{Vladilo}) for larger \hone\
column densities.
The conclusion again is that this system, although made of several individual
absorbing clouds, can be treated as one single cloud exposed to the
ionizing radiation.
This is basically the same
conclusion reached by 
Prochaska \& Wolfe (\cite{Prochaska2}) for the $z=2.5$ DLA. Let us
also note that a constancy of the ionization conditions 
is similarly observed in the local ISM through the small variability
of ionized helium 
(Wolff, Koester \& Lallement~\cite{Wolff}). 

\subsection{Spatial distribution of \althree}

\begin{figure}
      \vspace{0cm}
\hspace{0cm}\epsfig{figure=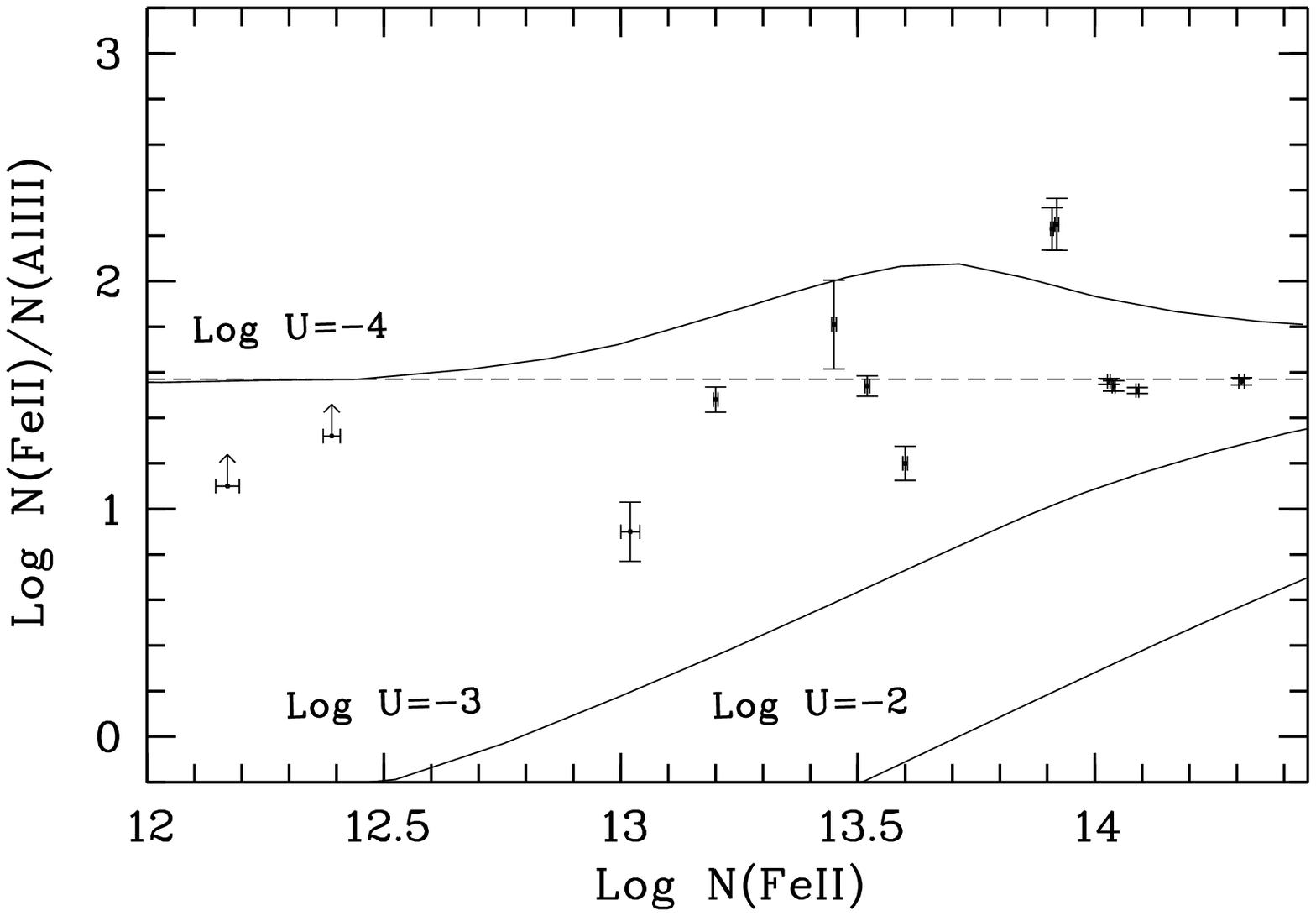,width=8.8cm}
\vspace{0cm} \caption[]{ Same as in Fig.~\ref{fig_ionization_a} but for 
the ratio $\log N(\fetwo)/N(\althree)$.
 } \label{fig_ionization_b}
      \end{figure}

Fig.~\ref{fig_ionization_b} also provides hints on the spatial
distribution of \althree. In view of our finding that the
low-ionization species in this DLA track each other independently of
gas density, if \althree\ is co-spatial with the low ions then we
expect that they anticorrelate, because a decrease in \hone\ (low-ion)
column density implies an increase in ionization. Since little
dependence of \althree\ with \fetwo\ is observed, we
conclude that a scenario of partially ionized clouds (i.e., with
\althree\ occuring in partially ionized interfaces bordering the
neutral gas) is  more plausible than co-spatiallity.

In such a scenario one would expect from the line widths to observe no
correlation between \althree\ and \altwo\ $b$-values. We obtain a
dispersion of $\sigma \equiv\sqrt{\sum{(\Delta b)^2}/11}=1.94$ \kms\
for \altwo--\althree, while $\sigma=1.1$ \kms\ for \altwo--\fetwo, and
$0.7$ \kms\ for \fetwo--\sitwo\ (including C13 and C14). If real, these
differing dispersions are indicative of absorption occuring in
different regions.\footnote{ We stress, however, the potentially large
systematic errors that may be present in our measurements of $b$,
especially due to saturation of some \altwo\ components.}

%
%
%

%

\subsection{Photoionization models}

We performed 
photoionization models for this DLA and compared the simulations with
the observed 
\sitwo, \fetwo, and \althree\ in order to establish the level of ionization of
each {\it single} cloud.  
The simulation outputs were also used to  quantify possible (if any)
ionization corrections to the observed relative abundances.   

We used the code CLOUDY (Ferland~\cite{Ferland}) to model the
ionization state for a range of \hone\ column densities and ionization
parameters ($U$, the ratio of ionizing photons to the total hydrogen
number density).  CLOUDY gives column densities for a given set of the
input parameters: gas metallicity $Z$, gas number density $n_{\rm H}$,
and ionizing spectrum $J_{\nu}$. Of these, shape and intensity of
the ionizing spectrum introduce the major uncertainties. We adopted
the universal background ionizing field due to QSOs and galaxies
calculated in Haardt \& Madau (\cite{Haardt}; an electronic, updated
version was kindly provided by the authors). This is the appropriate
choice for our treatment of the individual clouds as isolated objects
embedded in an intergalactic field.  Stars as sources of
photoionization are not likely to influence significantly  these
models: again referring to the work by Prochaska \& Wolfe
(\cite{Prochaska2}) 
the exponential fall off of a blackbody ionizing flux has a
similar shape as the \hetwo\ Lyman-break in the Haardt \& Madau
spectrum for the energy 
range of interest. The intensity of the field at the
Lyman-limit was taken to be $J_{912}=4.4\times 10^{-22}$
ergs~s$^{-1}$cm$^{-2}$Hz$^{-1}$sr$^{-1}$ at $z=2.3$ (consistent with
results obtained using the proximity effect; e.g.,
Giallongo~\cite{Giallongo}). The geometry was plane parallel and 
illuminated from one side.  The metallicity was fixed at [Fe/H]
$=-1.0$, and the relative abundances of interest [Si/Fe] and [Fe/Al] were
fixed at the values listed in Table~\ref{tbl-5}.  For fixed values of
$U$ we ran a grid of 
models by varying $N(\hone)$ in steps of $0.2$ dex. Each integration
was stopped when the required \hone\ value was reached. 
(i.e., the cloud size scales with
\hone\ in order to keep $U$ constant). The smoothed curves in
figures~\ref{fig_ionization_a} and~\ref{fig_ionization_b} show the
results of the simulations for $\log U=-2$, $-3$, and $-4$.

Fig.~\ref{fig_ionization_a} shows that in our model even at the very
lowest ionization parameters a mild but significant increase of the
\sitwo/\fetwo\ ratio with decreasing \fetwo\ is expected. Compared to
the model, the datapoints
confirm that the behavior of the
observed $\sitwo/\fetwo$ ratio is not consistent with isolated clouds,
and that some degree of self-shielding must exist in order to keep
this ratio constant over the observed range of \fetwo\ column
densities. We warn that the  \hone\ value at which the iterations are
stopped to obtain a given 
(\fetwo,\sitwo/\fetwo)-point is systematically lower than  
our assumed $\log N(\hone)=\log N(\fetwo) + 5.75$. In
other words, \fetwo\ is not tied to \hone\ in this model for
$N(\fetwo) <\sim 13.5$ \icm. The simulated \fetwo\ values are indeed 
obtained for systematically higher \hone\ column densities than the
assumed from the observed \fetwo\ values. However, this fact and the use in the
simulations of slabs illuminated just from one side both support the
notion that {\it i)} the individual clouds are embedded in a more neutral
environment than if they were isolated, and {\it ii)} the ionization is
homogeneous over the DLA components. In consequence, treating the
system as one single cloud with $N(\hone)=20.67$ should not introduce
major systematic errors.

Concerning \althree,  note that all but 2 points in
Fig.~\ref{fig_ionization_b} lie  
within the contours $\log U=-4$ and $-3$, while all \sitwo\
measurements are explained with $\log U <-4$. If different
ionization conditions explain the data, then \althree\ and \sitwo\
(and so \altwo) may occur in spatially distinct regions.

\begin{figure}
      \vspace{0cm}
\hspace{0cm}\epsfig{figure=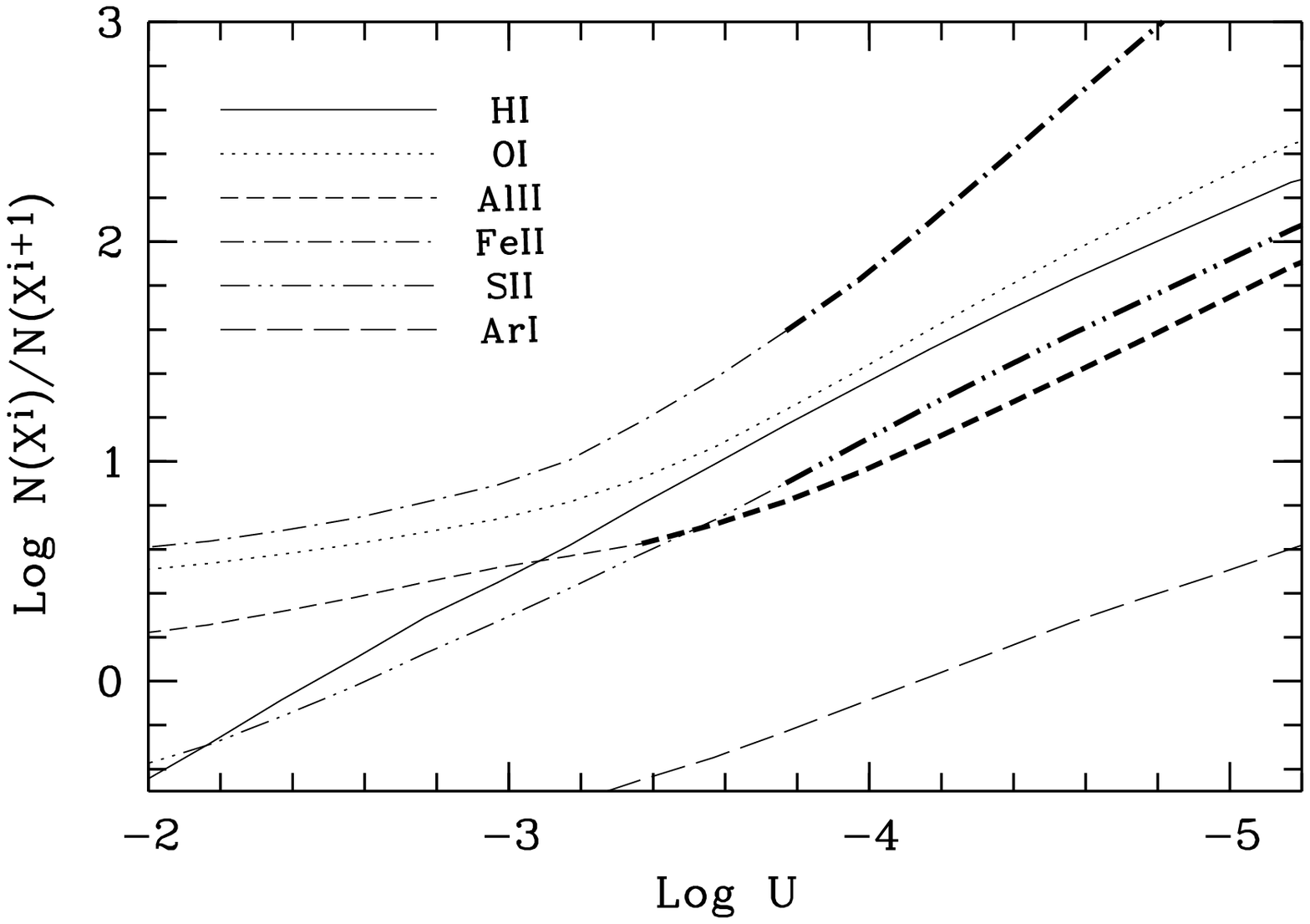,width=8.8cm}
\vspace{0cm} \caption[]{ Expected column density ratios as a function of
ionizing parameter $U$ for $\log
N(\hone)=20.67$ and a Haardt \& Madau (\cite{Haardt}) ionizing field.

 } \label{fig_ionization_DLA}
      \end{figure}

Finally, Fig.~\ref{fig_ionization_DLA} considers the same
photoionization models but with $N(\hone)=20.67$. The vertical axis
now shows the ratio $N({\rm X}^i)/N({\rm X}^{i+1})$ for the variety of
observed ions as a function of U. The thick sections of the curves
indicate those regions 
allowed by the $3\sigma$ upper limits on the doubly ionized
species \althree, \fethree\ and \sthree. The softer limit is set by
\altwo/\althree, but recall that 
\altwo\ is probably underestimated due to mild saturation. {\bf Assuming
$\log U\approx -3.8$, this plot
shows the approximation [X/H] $\approx$ [X$^{+}$/H$^0$] should be
accurate for the present DLA to within $\sim 5$ \%,} with the
exception of Ar, which is still primarily ionized in this regime.

%

\section{Metal abundances}
\label{abundances}


   \begin{table}
      \caption[]{Low ions: Total column densities and abundances}
         \label{tbl-5}
      \[
         \begin{tabular}{lcccc}
            \hline
            \noalign{\smallskip}
{\rm Ion}&$f^a$&$\log N_{\rm fit}^b$&$\log N_{\rm corr}^c$&{\rm [X/H]$^d$}\\
            \noalign{\smallskip}
            \hline
            \noalign{\smallskip}

	             			  	
\hone\ 1215     &  0.41640&  20.67(0.02)  &		  &           \\
\ctwo\ 1036     &  0.12310&$>$15.02(0.06) &$>$15.80   & $>$-1.43 0.08  \\
\ctwo\ 1334     &  0.12780&               &           &                \\
\none\ 1199     &  0.13280&  14.76(0.003) & 14.88     &    -1.84 0.02 \\    
\none\ 1200.2   &  0.08849&               &           &		      \\
\none\ 1200.7   &  0.04423&               &           &		      \\
\none\ 1134.1   &  0.01342&               &           &                \\
\none\ 1134.4   &  0.02683&               &           &                \\
\none\ 1134.9   &  0.04023&               &           &                \\
\oone\ 1039     &  0.00904&  16.21(0.19)  & 16.79     &    -0.81 0.21  \\    
\oone\ 1302     &  0.04887&               &           & 	  \\
\oone\ 950      &  0.00157&               &           & 	      \\
\oone\ 948      &  0.00645&               &           &                \\
\altwo\ 1670    &  1.81000&$>$13.83(0.02) &$>$13.97   & $>$-1.18 0.04  \\
\sitwo\ 1808    &  0.00256&  15.36(0.02)  & 15.36     &    -0.86 0.04  \\
\sitwo\ 1526    &  0.12700&               &           &		   \\
\ptwo\ 1152     &  0.23600&  13.28(0.08)  & 13.42     &    -0.82 0.10  \\
\ptwo\ 963      &  1.46000&               &           &		   \\
\stwo\ 1250     &  0.00545&  14.88(0.01)  & 15.02     &    -0.92 0.03  \\
\stwo\ 1253     &  0.01088&               &           &		   \\
\stwo\ 1259     &  0.01624&               &           &		   \\
\arone\ 1048    &  0.24400&  13.57(0.05)  & 13.85     &    -1.38 0.07  \\
\crtwo\ 2056    &  0.10500&  13.24(0.02)$^e$ &        &    -1.11 0.09  \\
\mntwo\ 2606    &  0.19270& $<$11.6       &$<12.43$   & $<$-1.77	   \\
\fetwo\ 1608    &  0.05850&  14.92(0.003) & 14.92     &    -1.26 0.02 \\
\fetwo\ 2260    &  0.00244&               &           &		   \\
\fetwo\ 2344    &  0.11420&               &           &		   \\
\fetwo\ 2374    &  0.03131&               &           &		   \\
\fetwo\ 2382    &  0.30000&               &           &                \\
\nitwo\ 1709    &  0.03240&  13.86(0.03)  & 14.00     &    -0.92 0.05  \\
\nitwo\ 1741    &  0.04270&               &           &		   \\
\zntwo\ 2026    &  0.48900&  12.22(0.03)$^e$  &       &    -1.10 0.05  \\
								   
            \noalign{\smallskip}
            \hline
         \end{tabular}
      \]
\begin{list}{}{}
\item[$^a$] {\small Oscillator strengths taken from 
Morton (\cite{Morton}) for \hone, \ctwo, \none, \ptwo,
\stwo, \arone, and \mntwo; 
from  Verner et al. (\cite{Verner}) (updated values) for \oone, \altwo,
\althree, and \sthree; from Schectman et al. (\cite{Schectman}) for
\sitwo $\lambda 1526$;
from Bergeson \& Lawler (\cite{Bergeson2}) for \sitwo $\lambda 1808$;
from Bergeson \& Lawler (\cite{Bergeson1}) for \crtwo\ and \zntwo;
from Fedchak et al. (\cite{Fedchak}) for \nitwo; 
from Howk et al. (\cite{Howk1}) for \fetwo $\lambda\lambda 
1608,2260,2344$, and $2374$; and 
from Cardelli \& Savage (\cite{Cardelli}) for \fetwo $\lambda 2382$. }
\item[$^b$] {\small Overall column densities from Voigt-profile
fits. For \ctwo, \oone, and \altwo\ at least one component was
saturated in all transitions available.}
\item[$^c$] {\small Column densities corrected for non-fitted
components using weights given by the individual $N(\fetwo)$-values.}
\item[$^d$] {\small [X$^{+}$/H$^{0}$], except for \none, \oone, and
\arone.}  
\item[$^e$] {\small Column density derived from the
apparent-optical-depth method.} 
\end{list}

   \end{table}

   \begin{table}
      \caption[]{High ions}
         \label{tbl-6}
      \[
         \begin{tabular}{lccc}
            \hline
            \noalign{\smallskip}
{\rm Ion}&$f^a$&$\log N_{\rm fit}$&$\log N_{\rm corr}^c$\\

            \noalign{\smallskip}
            \hline
            \noalign{\smallskip}


\althree\ 1862  &  0.28600&  13.35(0.01)$^b$  & 13.35    	   \\
\althree\ 1854  &  0.57500&               &          	   \\
\sthree\ 1012   &  0.04250&$<$14.05$^d$   &$<$14.05   	   \\
\fethree\ 1122  &  0.07884&$<$13.40$^e$   &$<$14.20   	   \\
\pthree\ 998    &  0.11200&$<$13.20$^f$   &$<$14.00   	   \\
\ntwo\ 1083     &  0.10310&$<$13.95$^g$   &$<$14.55   	   \\
            \noalign{\smallskip}
            \hline
         \end{tabular}
      \]
\begin{list}{}{}
\item[$^a$] {\small Oscillator strengths taken from 
Morton (\cite{Morton}) for \fethree\ and \ntwo; 
from  Verner et al. (\cite{Verner}) for \sthree, \pthree\ and \althree.}
\item[$^b$] {\small Overall column density from Voigt-profile
fits.}
\item[$^c$] {\small Column densities corrected for non-fitted
components using weights given by the individual $N(\althree)$-values}
\item[$^d$] {\small Column density limit derived from the
apparent-optical-depth method} 
\item[$^e$] {\small Column density limit derived from Voigt profile of
C3 assuming $b_{\fetwo}=b_{\fethree}$}  
\item[$^f$] {\small Column density limit derived from Voigt profile of
C3 assuming $b_{\ptwo}=b_{\pthree}$} 
\item[$^g$] {\small Column density limit derived from Voigt profile of
C11 assuming $b_{\none}=b_{\ntwo}$} 
\end{list}

   \end{table}

\subsection{Corrections to column densities}
\label{corr}

For all ions but \fetwo\ and \sitwo, we were not able to determine column
densities of each of the  14 velocity components.  For weak transitions in
noisy parts of the spectrum only the stronger components could be
fitted, while for strong transitions only the weakest components where
accessible due to blending of the saturated components. For those
incomplete fits we applied a correction by scaling the summed column
densities according to the results for \fetwo.  We have shown in the
previous section that this approach is acceptable in this 
DLA. The corrected values are listed in the column labeled $N_{\rm
corr}$ of Table~\ref{tbl-5}. 
Since for weak transitions the strongest components dominate the
total column densities, the corrections are small and should not
introduce large uncertainties; on the contrary, intrinsically strong
transitions with only weak components fitted (\oone\ and \ctwo),
should suffer from systematic errors, and must thus be taken
cautiously.

\subsection{Metallicity of the DLA gas}

Taking the non-refractory element Zn as reference, the gas metallicity
for this DLA {\bf turns out to be [Zn/H] $=-1.10\pm 0.05$ (or $Z_{\rm
DLA}=1/12~Z_{\sun}$). A slightly lower metallicity is obtained  
if Fe is used,} [Fe/H]
$=-1.26\pm 0.02$. For $z=2$, this  Fe abundance 
is quite at the upper end of current samples of DLA systems 
(Prochaska \& Wolfe~\cite{Prochaska1}), while the Zn value is much
more representative of all measurements at $z\sim 2-2.5$ ((Pettini et
al.~\cite{Pettini}; Vladilo et al.~\cite{Vladilo2}). This is because
[Zn/Fe] $> 0$ for the majority of the DLAs.

\subsection{Dust content}
\label{dust}

The key property of the present DLA system is its apparent lack of
dust. 
{\bf We measure [Zn/Cr]
$=-0.01\pm0.05$,} a value which is consistent with no dust-depletion,
since Cr, unlike Zn, is depleted in the local ISM.
Another check for the presence of dust is provided by the refractory
elements  Fe
and  Ni. The Zn/Fe and Zn/Ni  ratios are expected to be supersolar if
dust is present, because Fe and Ni are two of the most heavily
depleted elements in the ISM. In addition, [Zn/Fe] $\approx 0$ for
metal-poor stars of a wide range of metallicities (Gratton \&
Sneden~\cite{Gratton}; Sneden, Gratton \& Crocker~\cite{Sneden}; but
see the discussion in Prochaska et al.~\cite{Prochaska4}), so any
departure from this value could be attributed to dust depletion.  {\bf
We
observe [Zn/Fe] $=+0.16\pm 0.03$ and [Zn/Ni] $=-0.18\pm 0.06$. 
While the Zn/Fe ratio implies a certain degree of dust depletion, the
Zn/Ni abundance gives further evidence in support of dust-free gas. }
Altogether these 3 ratios
indicate that the fraction of atoms that are missing from the gas
phase is insignificant.  

Nevertheless, these inferences all rely on the goodness of our zinc
abundance, {\bf which was not determined by profile fitting of the velocity
components. 
An independent argument against significant dust depletion comes from
the ratio [S/Si] $-0.06\pm0.03$.} Since (a) these two 
elements have a common nucleosynthetic origin, and (b) silicon is
normally depleted in the ISM while sulfur is not (Savage \&
Sembach~\cite{Savage}), one expects this ratio to be supersolar if
dust were present in this DLA, which  is not observed.

In conclusion, it seems that we have found a new DLA lacking dust, the
second such instance at high redshift. 
Other cases of
DLAs with small dust content are reported in Molaro et
al. (\cite{Molaro1}), in Pettini et al. (\cite{Pettini2}), and in
Prochaska \& Wolfe (\cite{Prochaska3}). From the Pettini et
al. compilation the DLA with lowest dust content has [Zn/Cr]
$=-0.13\pm 0.20$ at $z=0.86$ toward Q0454+039, and from Prochaska \&
Wolfe, [Zn/Cr] $=-0.20\pm 0.12$ in the $z=1.920$ DLA toward
Q2206-199. In both cases, there appears to be some degree of dust
depletion, at least stronger than in the present DLA. A more similar
case as ours was reported by Molaro et al. (\cite{Molaro1}) who
measured [Zn/Cr] $=-0.06\pm 0.08$ at $z=3.39$ toward Q0000-2620. 
{\bf This shows that  the present
system is uncommon among the few documented low-dust DLAs.}  

%
%



\subsection{$\alpha$/Fe-peak ratios}

A diagnostic tool to probe the evolutionary stage in DLAs 
is the abundance ratio of $\alpha$ to Fe-peak elements. Due to
their different origins, $\alpha$-capture elements being produced
mainly in SNe Type II, and  Fe-peak elements being delivered to the ISM
by SNe Type Ia on a longer timescale, the $\alpha$ to Fe-peak ratio is
expected to rapidly attain supersolar values in a chemically less
mature  system.
Moreover, this ratio should decrease as cosmic time proceeds, unless
possible environmental factors might influence it (Ellison \&
Lopez~\cite{Ellison}). 
In our Galaxy the values range from $<[\alpha/{\rm
Fe]}>$ $\approx +0.5$ in metal-poor stars down to $\approx 0$, as
metallicity increases.

All $\alpha$ elements we measure in the DLA toward \he\ follow each
other quite closely (with the exception of Ar; see below). 
{\bf We obtain [Si/Zn] $=+0.24\pm0.05$, and [S/Zn]
$=+0.18\pm 0.04$ }(recall that silicon is normally depleted in the local
ISM; Savage \& Sembach~\cite{Savage}). 
 Even the $\alpha$-element oxygen, despite the
larger measurement errors, fits well into the trend {\bf with [O/Zn]
$= +0.29\pm0.22$. On average, we obtain for Si and S  $<[\alpha/{\rm
Zn}]>= +0.21\pm 0.06$ }(simple mean). 
This value is in good
agreement with measurements  of metal-poor halo stars (e.g., Ryan et
al.~\cite{Ryan} and references therein) or the thick disk (Prochaska
et al.~\cite{Prochaska4}) at [Fe/H] $=-1.2$, but is high when
compared with the current sample of DLAs.
The conclusion is: like 
in Galactic metal-poor stars, these ratios probably reflect the temporal delay
between SNe of Type I and II. They show that, despite its high
metallicity, this DLA gas has not undergone significant metal pollution
from SN Type I yields. We re-emphasize that these ratios are {\bf
likely} unbiased 
from dust-depletion  effects. Centurion et al. (\cite{Centurion}) have used
the [S/Zn] ratio to investigate a possible evolution of the
$\alpha$/Fe-peak abundance ratio that is unbiased from dust
effects. Although their sample is small (6 measurements) those authors
do find a decrease in that ratio with 
increasing metallicity, which is what one would expect if metallicity
keeps track of  chemical evolution in DLAs and if the
objects conform a chemically  homogeneous sample. Our measure of
[S/Zn] does not fit in that trend, being too high for [Zn/H]
$=-1.1$, but it much better conforms to the enhancement observed in
metal-poor stars.


\subsection{The odd-even effect}

The odd-even effect, that is, the underabundance of odd-$Z$ elements
relative to even-$Z$ elements of the same nucleosynthetic origin, is
another known property of Halo stars. Since Fe is more prone to  dust
depletion than Mn in a variety of ISM environments, this ratio can be
used as a discriminant between dust depletion and pure SN Type II
enrichment. In fact, [Mn/Fe] averages $\approx -0.3$ in Galactic Halo stars
at [Fe/H] $=-1$ (Ryan et al.~\cite{Ryan}), whilst [Mn/Fe]
$\approx +0.3$ in the local ISM (Savage \& Sembach~\cite{Savage}).

The \fetwo -corrected value of our $3\sigma$ upper limit for the non-detection
 of \mntwo\ yields [Mn/Fe] $< -0.51$, which confirms the odd-even
 effect, is in accordance with the sample of Halo metal-poor stars in
 Ryan et al. (\cite{Ryan}), but is quite low for that of thick disk
 stars in Prochaska et al. (\cite{Prochaska4}). The low value also
 conforms with previous DLA measurements (Lu et al.~\cite{Lu}).

We do not observe the same effect for P, the other odd-element
observed in this DLA. {\bf We obtain [P/Si] $=+0.10\pm 0.09$ at [P/H]
$=-0.8$. This value is high if compared with  [P/Si] $=-0.40$ at
[P/H] $=-2.3$ (Molaro et al.~\cite{Molaro}) and [P/Si] $=-0.30\pm
0.09$ at [P/H] $=-1.2$ (Outram et al.~\cite{Outram}).  }

\subsection{Underabundant \arone}

Our column density estimate for \arone\ implies a significant
underabundance of Ar with respect to other $\alpha$-chain
elements. This corresponds to $\sim -0.5$ dex of what is expected if
argon {\bf is overabundant by $+0.2$ dex} (assuming argon tracks $\alpha$
elements [Timmes, Woosley \& Weaver~\cite{Timmes}]). In the local ISM,
\arone\ is rarely found or is 
significantly below its cosmic abundance relative to \hone\ (Sofia \&
Jenkins~\cite{Sofia}; Jenkins et al.~\cite{Jenkins}). 
As suggested by Sofia \& Jenkins in order to
explain their finding of a low \arone\ abundance in the local ISM, the 
underabundance can be explained if a significant part of the argon is
ionized. In this    
DLA that fraction is $\sim 60$ \%.  Moreover, since \arone\ has a much larger
ionization cross section than \hone, our measurement can also be
explained in terms of moderately ionized gas, which brings \arone\ into
\artwo\ in higher proportions than for the rest of the low-ions.  The
alternative explanation, a low \arone\ abundance caused by dust
depletion, is in our DLA ruled out given (1) the scarce  propensity of Ar
to dust-grain incorporation (Meyer et al.~\cite{Meyer}; Sofia \&
Jenkins~\cite{Sofia})  and (2) the negligible amount of dust
(\ref{dust}). Thus, these conditions resemble very much those in the
local ISM described by Sofia \& Jenkins where argon is systematically
found underabundant, and confirm their suggestion that ionization, and
not dust, is the responsible agent.

\subsection{Nitrogen}

Chemical evolution models predict the production of nitrogen in the
CNO cycle in stellar interiors to have two different
origins. While primary N is produced in intermediate-mass stars (3 --
8 $M_{\sun}$) without dependence on the initial metal content,
secondary nitrogen is also produced in stars of any mass but with an
initial supply of heavy elements. The net result is that the N/O ratio
should be independent of O/H for primary N while N/O $\propto$ O/H for
secondary N. In addition, since intermediate-mass stars evolve on much
larger scales than the short-lived progenitors of SN Type II -- responsible for
the bulk of $\alpha$-elements like O, Si, S, etc.  --, there should be
a delay of $\sim 5\times 10^8$ yr 
between the release of primary N and that of O (Lu et al.~\cite{Lu1}
and references therein).

We measure [N/O] $=-1.03$ at [O/H] $=-0.81$. However, 
despite the accurate N abundance we were able to obtain, 
our oxygen estimate is rather
uncertain.\footnote{Fig.~\ref{fig_variations} shows C8 departures from the
trend followed 
by the other 4 components. This might be due to an overestimate of C8
(which dominates the total column density).  Excluding C8 and
re-scaling, however, implies lowering the total column density by
$0.7$ dex. Although such value conforms better with the rest of the
components, it translates into [O/H]
$=-1.5$, so the underabundance of N over O would become only [N/O]
$=-0.2$.}
Relying rather on the better estimates for Si and S, we find [N/Si] $=-0.98$
{\bf at [Si/H] $=-0.86$ (and basically the same values are found if S }
is used). Referring to Fig. 1 in Lu et al. (\cite{Lu1}), this value
is at the very low end of DLA measurements at this Si abundance;
moreover, according to the fits to Galactic measurements, it is better
explained as secondary nitrogen\footnote{{\bf But let us note that \ntwo\ 
could contribute with $\approx 0.20$ dex to the nitrogen abundance,
thus  bringing [N/Si] more in line with the Lu et 
al. values.}}. This is in agreement with the 
high $\alpha$/Fe-peak ratios we observe because both effects can be
explained by an early phase in this protogalaxy's chemical evolution, 
observed at a time when  the release of Fe-peak elements and primary
nitrogen has not yet begun.




\section{Summary of the conclusions}
\label{conclusions}

\begin{enumerate}

\item{We have studied in detail a $z=2.33$ DLA observed at high
resolution toward the bright  QSO \he. We have been  able to detect and
obtain reliable column densities for 
\hone, \none, \sitwo, \ptwo, \stwo, \arone,
\crtwo, \fetwo, \nitwo, \zntwo, and \althree; less accurate
determinations for \ctwo, \altwo\ 
and \oone; and to put upper limits
on \mntwo, \sthree, \fethree, \pthree, and \ntwo. {\bf The metallicity of this
system is [Zn/H] $= -1.1$.}
}

\item{By comparing column densities of single velocity components we
have performed the first cloud-by-cloud analysis in a high-$z$ DLA so
far. We find a tight correlation of all low-ionization species with
respect to \fetwo\ extending over 2.5 orders of magnitude in
$N(\fetwo)$. We interpret this trend as a lack of significant {\bf changes
in the} dust, ionization, and intrinsic chemical-enrichment effects. If
\fetwo\ tracks \hone, such homogeneity differs from what is observed
in the local ISM, meaning that our line of sight probably samples
clouds located in one single environment.

}

\item{We find a probable scenario for the occurrence of \althree\ is
ionized shells  surrounding neutral cores of \altwo. Photoionization
models show 
that the ratio of \althree\ to low ions can be modeled with a wider
range of higher ionization parameters than the ratio among low
ions. Independently of this, the disparate line width differences between
\altwo\ and \althree\ also support the notion of separate \altwo\ and
\althree\ regions.

}

\item{

Using different abundance ratios between ISM refractory and
non-refractory elements we have shown that the observations are 
compatible with negligible amounts of dust in this protogalaxy,
{\bf although the measured [Zn/Fe] $= +0.16\pm0.03$ may indicate a small
degree of dust depletion.}

}

\item{We observe -- unbiased by dust depletion effects -- 
an enhancement of the $\alpha$/Fe-peak {\bf ratios of $+0.2$
dex} for various elements, a marked odd-even effect in Mn, and a strong
underabundance of N relative to Si and S, [N/Si,S] $=-1$ at [Si/H]
$=-0.86$. All of these ratios support an environment that is in an early
evolutionary stage, where the onset star formation has begun shortly
before the DLA was observed. The absence of dust is in agreement with
this notion, if there has not been time enough to form dust grains. 

}

\item{The observed underabundance of \arone\ by $0.5$ dex is easily
explained by a significant fraction of argon residing in a higher ionization
phase. This situation resembles the conditions in the local ISM.
Our ionization simulations demonstrate, however, such an ionized phase does
not contain a considerable part of the other elements, so our
abundance measurements are not affected by undetected ionized atoms.

}

\end{enumerate}

\subsection{Final comment}

The observed uniformity of metal abundances across the line of sight
to \he\ has important implications for current models of galaxy formation.
As mentioned before, it suggests the individual components are  embedded
in the same environment, which, in addition, must have
undergone significant mixing. 
If DLAs are formed from the accretion of protogalactic small clumps
(e.g., Haehnelt et al.~\cite{Haehnelt}) that have formed
independently, then one expects each protogalaxy to  
contribute with gas of different dust and nucleosynthetic
compositions. The
homogeneity of metal abundances observed in this DLA is difficult
to explain in this scenario, unless
the mixing of metals has occurred in short enough timescales to
counteract the enhancement of any original differences (e.g., through
the metal release by  SN Type I; see the discussion in Prochaska \&
Wolfe~\cite{Prochaska5}). The chemically young gas we apparently
observe might be a 
consequence of such a rapid process. 

Although the statistical weight of our result on single-cloud
abundances may not be sufficient to draw any general conclusions on
the physical nature of DLAs, the observed uniformity in \he\ seems to
be a quite common property among DLAs (Prochaska et
al.~\cite{Prochaska6}). 

On larger physical scales, also important for constraining the models 
is to investigate abundance variations among systems.  There are few known
`DLA groups' along single lines of sight where the redshift separation is
small enough to support a physical connection between the DLAs, but
big enough to characterize the systems as separate entities (i.e., a
few $1\,000$ \kms; Lopez et al.~{\cite{Lopez1}; Ellison \&
Lopez~\cite{Ellison}). In the only such groups so far investigated at
high resolution, Ellison \& Lopez
(\cite{Ellison}) find very similar relative abundances among the DLA
members, despite their different metallicities. While it is not clear
whether such DLA groups later merge, environment may be an effective
catalyzer of metal enrichment in DLAs at early evolutionary stages.

\begin{acknowledgements}

We warmly thank Sara Ellison for her valuable
comments on an earlier version of the paper; Ulysses Sofia for an
informative discussion; Piero Madau and Francesco Haardt for
having made available an updated version of their spectrum to us; 
Scott Burles for having taken the B\&C spectrum of \he, and
G. Vladilo, the referee, for a careful reading of the 
manuscript. JXP was supported by NASA through a Hubble Fellowship grant 
HF-01142.01-A awarded by  STScI, and  SL acknowledges financial
support by FONDECYT 
grant N$^{\rm o} 3\,000\,001$ and by the Deutsche Zentralstelle f\"ur
Arbeitsvermittlung.

\end{acknowledgements}

\end{document}